\documentclass{article}
\usepackage[utf8]{inputenc}
\usepackage[margin=1in]{geometry}
\usepackage{amsmath}
\usepackage{physics}
\usepackage{graphicx}
\usepackage{amssymb}
\usepackage{simpler-wick}
\usepackage{authblk}

\newcommand{\q}{q}
\newcommand{\bbeta}{\mathfrak b}
\newcommand{\SFF}{\text{SFF}}
\newcommand{\Zs}{Z_{\textrm{enh}}}
\newcommand{\trp}{\textrm{TRP}}

\begin{document}
\title{Emergent Spectral Form Factors in Sonic Systems}
\author[1]{Michael Winer}
\author[2]{Brian Swingle}
\affil[1]{Joint Quantum Institute, Department of Physics, University
of Maryland, College Park, Maryland 20742, USA}
\affil[2]{Department of Physics, Brandeis University, Waltham, Massachusetts 02453, USA}
\maketitle 
\begin{abstract}
    We study the spectral form factor (SFF) for hydrodynamic systems with a sound pole, a large class including any fluid with momentum conservation and energy conservation, or any extended system with spontaneously broken continuous symmetry. We study such systems in a finite volume cavity and find that the logarithm of the hydrodynamic enhancement to the SFF is closely related to the spectral form factor of a quantum particle moving in the selfsame cavity. Depending upon the dimensionality and nature of the effective single-particle physics, these systems exhibit a range of behaviors including an intricate resonance phenomenon, emergent integrability in the SFF, and anomalously large fluctuations of the SFF.
    
\end{abstract}
\section{Introduction}
The statistics of the energy spectrum is one of the most important diagnostics of quantum chaos~\cite{haake2010quantum,PhysRevLett.52.1,mehta2004random}. There is strong evidence that ensembles of chaotic systems have the same Hamiltonian spectral statistics as ensembles of Gaussian random matrices, with examples including nuclear systems~\cite{doi:10.1063/1.1703775,wigner1959group}, mechanical systems \cite{berry1977level,McDonald,BERRY1981163}, condensed matter systems~\cite{bohigas1984chaotic,altshuler1986repulsion,andreev1995spectral,dubertrand2016spectral,PhysRevLett.121.264101,Wittmann_W__2022,speck,bunin2023fisher,kamenev_1999,PhysRevResearch.3.L012019}, holographic models~\cite{Cotler2017,Saad:2018bqo,PhysRevD.98.086026,saad2019late,saad2021wormholes,saad2022convergent} and (generalizing to time-dependent Hamiltonians) circuit models \cite{shivam_2023,Chan_2018,Chan_2018Min,Chan_2019,Liao_2022}. A central quantity in spectral statistics in the Spectral Form Factor (SFF). Among other interesting properties \cite{PhysRevLett.127.230602,CCRM}, this quantity diagnoses whether energy levels repel as they do in random matrices~\cite{bohigas1984chaotic,Kos_2018,Flack_2020}, have independent Poissonian statistics~\cite{berry1977level}, or have some more exotic behavior~\cite{2020Prosen,PhysRevLett.125.250601,Winer_2020,Li_2021}. The SFF has spawned relatives including the partial SFF \cite{Joshi_2022}, the entanglement SFF \cite{ma2022quantum,2021,shaffer_2014}, and the Loschmidt SFF \cite{WinerLoschmidt,Weidenmuller_2005,guhr1998random}.

The SFF can be written as
\begin{equation}
    \SFF(T,f) =\overline{ | \text{Tr}[U(T) f(H)]|^2},
\end{equation}
where $U(T) = e^{-i H T}$ is the time evolution operator, $f$ is a filter function (perhaps a Gaussian centered on a specific energy band), and overline denotes a disorder average over an ensemble of Hamiltonians. These Hamiltonians can differ either through microscopic disorder or, as we will see, through the overall shape of the region containing the system. For a discussion of how different the Hamiltonians must be to constitute ``sufficient coverage'', see~\cite{WinerLoschmidt,Weidenmuller_2005,guhr1998random,https://doi.org/10.48550/arxiv.2205.12968,PhysRevLett.70.4063}. This disorder average is necessary because while the disorder-averaged SFF has smooth behavior as seen below, the SFF of a single Hamiltonian is a highly erratic function of time. A central problem in the field of quantum chaos is thus to calculate the SFF for a variety of physical systems in order to understand the general conditions under which random matrix behavior emerges.

In ~\cite{WinerHydro,Winer_2022}, the authors set forth a hydrodynamic theory of the spectral form factor. This effective theory predicts pure random matrix behavior at late time and computes corrections at earlier times due to slow modes. Intuitively, the presence of slow modes is related to atypically small matrix elements in the Hamiltonian, and atypically small matrix elements suppress the level repulsion which is characteristic of a random matrix. These corrections in the SFF persist until the system's ``Thouless time'', after which the random matrix behavior is recovered. In particular, approximate symmetries enhance the ramp by an amount exponential in system size, consistent with \cite{Chan_2018,Friedman_2019,moudgalya2020spectral,kos_2021} . This enhancement factor was calculated for simple models of a system with conserved modes and with spontaneous symmetry breaking. These calculations included tree-level effects and diagrammatic corrections not seen in traditional hydrodynamics. \cite{WinerHydro} dealt exclusively with theories first order in time, and \cite{Winer_2022} dealt with systems with no spatial extent. However, the problem of hydrodynamic systems with sound poles was left unstudied. In this work, we fill in this gap, pointing out that sound pole effects can lead to new phenomena such as exponentially increasing SFFs and spectral form factors with sensitive qualitative dependence on the precise shape of the system.

\subsection{Summary of Results and Sketch of Paper}
\label{subsec:SumSketch}
The theory in \cite{WinerHydro,Winer_2022} expressed the enhancement of the SFF in linear hydrodynamics as a product over all modes. Here hydrodynamics (hydro) refers to a effective theory of a system's long-wavelength and long-time dynamics which incorporates the effects of conservation laws and other general constraints while coarse-graining over physics at more microscopic scales. \cite{WinerHydro} considered only cases where the hydro equations were first-order in time while \cite{Winer_2022} only studied systems with no spatial extent. As such, this is the first work to use the methods of hydrodynamics to study SFFs in systems with sound poles. This framework is meant to apply to systems for which the underlying microscopic theory is quantum chaotic and the long-wavelength physics is described by oscillatory sound modes. We treat the sound modes at the quadratic level within hydrodynamics (except for an appendix discussing some effects of hydrodynamic interactions), but this quadratic approximation still allows for the the modes to decay and is consistent with an underlying microscopic chaotic dynamics.

For the SFF enhancement, our hydrodynamic theory predicts that an oscillatory mode with angular frequency $\omega$ decaying at rate $\lambda$ contributes to the enhancement like $\left|\frac{1}{1-e^{i\omega T-\lambda T}}\right|^2$. This expression, which is obtained in Section 2, rapidly oscillates with angular frequency $\omega$, and the product of many such terms can generate much richer time-dependence than is possible with purely decaying modes (such as diffusive modes). The work of Sections 3, 4, and 5 is dedicated entirely to the study of these products. For this purpose, the spectrum of allowed $\omega$s is very important, and we find a variety of different behaviors depending on the statistical properties of the $\omega$s. This spectrum is determined by Laplacian of the region, or ``cavity'', in which the system resides. In particular, it is crucial to understand whether the spectrum of sound modes admits constructive or destructive interference in the product over $\omega$s.%, i.e. whether the terms $\left|\frac{1}{1-e^{i\omega T-\lambda T}}\right|^2$ for different $\omega$ tend to be large at the same time.

In the case of 1d linear hydrodynamics, we find that patterns of interference impart an intricate fractal structure on the graph of the spectral form factor, as seen in figure~\ref{fig:babylon}. There is an exponentially tall peak at every rational time in units of the system length. Unfortunately, this result is delicate; at the very least, hydrodynamic interactions destroy the sound pole in 1d~\cite{spohn_2020}, with the dynamics flowing to to the KPZ universality class \cite{KPZ,Krug1997OriginsOS,krajnik_2020}. However, we still include this discussion as a particularly simple example of the dramatic effects of oscillations and because the sound pole can survive in some special situations, such as when the local Hilbert space dimension is large.

More generally, for cavities in higher dimensions, one sees an enhancement factor that is qualitatively exponential in the single-particle SFF for a billiard of the same shape as the cavity. This surprising statement means that even though the fundamental system is made out of many microscopic degrees of freedom undergoing aperiodic motion, the full system SFF depends very strongly on the motion of wavepackets of sound. We distinguish between ``integrable'' cavities, in which the spectrum of the Laplacian is Poissonian, and ``chaotic'' cavities in which the spectrum of the Laplacian is random matrix like (although we stress that this integrable or chaotic adjective refers to the structure of the cavity modes, not to the many-body levels, which are always ultimately random-matrix-like in our models).

For example, local disordered systems in integrable cavities such as tori or ellipsoids see an exponential (in volume) enhancement to the many-body SFF corresponding to the Poissonian statistics of sound wavepackets, whereas disordered systems in a chaotic cavity, such a two-dimensional region in the shape a Bunimovich stadium billiard, see an exponentially growing enhancement to their SFF corresponding to the ramp in the SFF of a sonic wavepacket.

The rest of the paper is organized as follows. In the remainder of the introduction, we briefly review the SFF. In Section 2, we review the hydrodynamic theory of \cite{WinerHydro} and discuss its application to systems with a sound pole. In Section 3, we treat a one-dimensional system where the SFF exhibits an intricate fractal structure. In Section 4, we move to higher dimensions and discuss the case of a sound pole when the system is confined to an integrable cavity. In Section 5, we discuss the analogous problem in a chaotic cavity. In Section 6, we discuss interaction effects, with the preceeding discussion all at the quadratic level. This is followed by a brief discussion of the results; appendix A also contains a comparison to our prior work on SYK2.

\subsection{Review of the form factor}
\label{subsec:SFFReview}
Before proceeding to the derivations of these results, we first review the SFF in more detail. The starting point for any understanding of the spectral form factor is level repulsion. If we take the well-known joint probability density function for the $N$ eigenvalues of an $N$ by $N$ random Hermitian (GUE) matrix, we have
\begin{equation}
\begin{split}
    P(\lambda_1,\lambda_2,\lambda_3...\lambda_N)=\frac{1}{Z_N}\exp(\frac{N}{2}\sum_i \lambda_i^2)\prod_{ij}(\lambda_i-\lambda_j)^2=\\
    \frac{1}{Z_N}\exp\left(\int_{\lambda_1>\lambda_2} 2\log (\lambda_1-\lambda_2)\rho(\lambda_1)\rho(\lambda_2)d\lambda_1d\lambda_2-\int \frac{N}{2}\lambda^2\rho(\lambda) d\lambda\right).
    \label{eq:JPDF}
\end{split}
\end{equation}
It is the last term in the product on the first line that is responsible for level repulsion. Intuitively, if we zoom in on a near-degeneracy involving two levels, then the $2$ by $2$ submatrix involving those two levels only has an exact degeneracy if the coefficients of each basis matrix (Pauli $\sigma_x$, $\sigma_y$, and $\sigma_z$) are all zero, so near-degeneracy requires a special degree of fine-tuning.

The SFF is closely related to the correlation function of the density of states.
Formally, the (filtered) density of states is given by
\begin{equation} \label{eq:general_density_of_states_definition}
\rho(E, f) \equiv \sum_n f(E_n) \delta(E - E_n) = \textrm{Tr} f(H) \delta(E - H),
\end{equation}
where $n$ labels the eigenstate of $H$ with eigenvalue $E_n$, and its correlation function is
\begin{equation} \label{eq:density_of_states_correlation_definition}
C(E, \omega, f) \equiv \mathbb{E} \left[ \rho \left( E + \frac{\omega}{2}, f \right) \rho \left( E - \frac{\omega}{2}, f \right) \right].
\end{equation}
We have that
\begin{equation} \label{eq:SFF_density_of_states_relationship}
\begin{aligned}
\textrm{SFF}(T, f) &= \mathbb{E} \Big[ \textrm{Tr} f(H) e^{-iHT} \textrm{Tr} f(H) e^{iHT} \Big] \\
&= \int dE d\omega \, e^{-i \omega T} \mathbb{E} \left[ \textrm{Tr} f(H) \delta \left( E + \frac{\omega}{2} - H \right) \textrm{Tr} f(H) \delta \left( E - \frac{\omega}{2} - H \right) \right] \\
&= \int d\omega \, e^{-i \omega T} \int dE \, C(E, \omega, f).
\end{aligned}
\end{equation}
The SFF is simply the Fourier transform of the correlation function with respect to $\omega$. 

The SFF can be split into two contributions:
\begin{equation} \label{eq:general_SFF_decomposed}
\textrm{SFF}(T, f) = \big| \mathbb{E} \textrm{Tr} f(H) e^{-iHT} \big|^2 + \bigg( \mathbb{E} \left[ \big| \textrm{Tr} f(H) e^{-iHT} \big|^2 \right] - \big| \mathbb{E} \textrm{Tr} f(H) e^{-iHT} \big|^2 \bigg).
\end{equation}
The first term, the disconnected part of the SFF, comes solely from the average density of states. It is the absolute value square of the density's Fourier Transform.

It is the second term, the connected part of the SFF, that contains interesting information on the correlation between energy densities.
``Random matrix universality''~\cite{PhysRevLett.52.1,Altland1997} is the principle that an ensemble of quantum chaotic Hamiltonians will generically have the same \textit{connected} SFF as the canonical Gaussian ensembles of random matrix theory~\cite{mehta2004random,tao2012topics}.
This behavior is illustrated in Fig.~\ref{fig:SFFgraph}, which plots the disorder-averaged SFF of the Gaussian unitary ensemble (one of the aforementioned canonical ensembles).
The graph shows the three regimes of the random matrix theory SFF:
\begin{figure}
\centering
\includegraphics[width=0.9\textwidth]{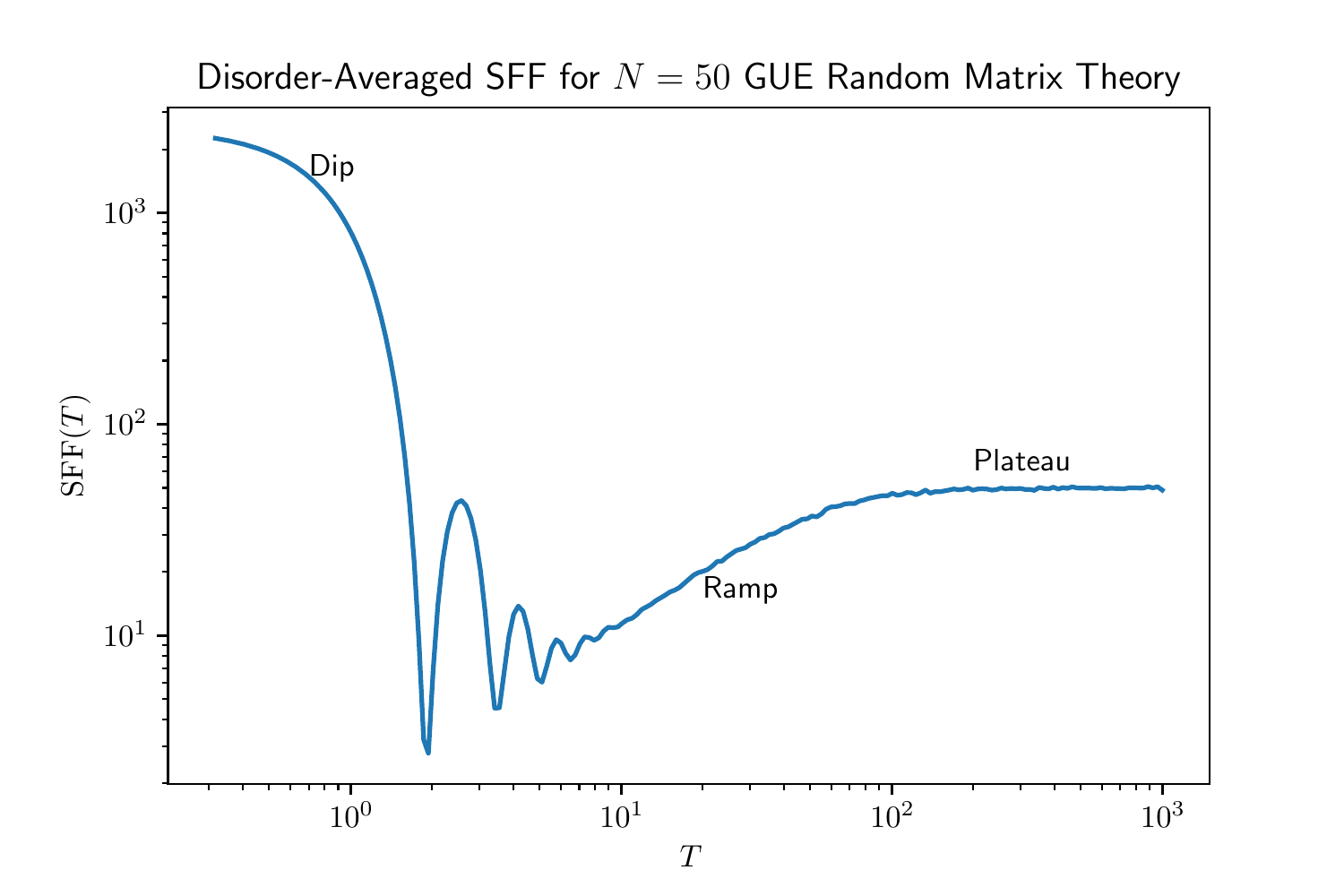}
\caption{The disorder-averaged SFF for the Gaussian unitary ensemble (GUE). The matrices in this ensemble have dimension $N = 50$. The SFF was computed numerically by averaging over ten thousand realizations. The three regimes --- dip, ramp, plateau --- are labeled.}
\label{fig:SFFgraph}
\end{figure}
\begin{itemize}
\item The ``dip'', occurring at early times, comes from the disconnected piece of the SFF (and thus its precise shape is non-universal and depends on the details of $f$ and the thermodynamics of the system).
Its downward nature reflects a loss of constructive interference --- the different terms of $\textrm{Tr} e^{-iHT}$ acquire different phase factors as $T$ increases.
\item The ``ramp'', occurring at intermediate times, is arguably the most interesting regime.
In canonical random matrix ensembles, the ramp follows from the result~\cite{mehta2004random}
\begin{equation} 
\mathbb{E} \left[ \rho \left( E + \frac{\omega}{2} \right) \rho \left( E - \frac{\omega}{2} \right) \right] - \mathbb{E} \left[ \rho \left( E + \frac{\omega}{2} \right) \right] \mathbb{E} \left[ \rho \left( E - \frac{\omega}{2} \right) \right] \sim -\frac{1}{\bbeta \pi^2 \omega^2},
\label{eq:repulsion}
\end{equation}
where $\bbeta = 1$, $2$, $4$ for the orthogonal, unitary, and symplectic ensembles respectively \cite{mehta2004random}.
The fact that the right hand side is negative is known as level repulsion in quantum chaotic systems \cite{wigner1959group}. Since the energy levels of a system repel each other, fluctuations in the density of states are suppressed. Long-wavelength (short time) fluctuations are suppressed the most, short-wavelength (long time) fluctuations are suppressed the least. 
Taking the Fourier transform of equation \ref{eq:repulsion} respect to $\omega$ gives a term proportional to $T$ for the connected SFF.
Such a linear-in-$T$ ramp is often taken as a defining signature of quantum chaos.
\item The ``plateau'', occurring at late times, is fundamentally a consequence of the discreteness of the spectrum.
At times much larger than the inverse level spacing or ``Heisenberg time'', all off-diagonal terms in the double-trace of the SFF average to zero, meaning that
\begin{equation} \label{eq:SFF_plateau_derivation}
\textrm{SFF}(T, f) \approx \sum_{mn} e^{-i(E_m - E_n)T} f(E_m) f(E_n) \sim \sum_n f(E_n)^2.
\end{equation}
For integrable systems, the plateau is reached very quickly with little to no ramp regime \cite{berry1977level,PhysRevLett.125.250601,Winer_2020}. For chaotic systems it isn't reached until a time exponential in system size. Nonetheless, it is the long-term fate of any system without a degenerate spectrum.
\end{itemize}
This paper will focus on the connected SFF at time scales towards the beginning of the ramp, scaling polynomially in system size instead of exponentially. At this time scale, physical systems do not behave exactly like random matrices, and exciting new phenomena can be observed. The timescale separating these non-universal phenomena from the linear ramp of RMT is known as the Thouless time, see e.g. \cite{Gharibyan_2018,Chan_2018} for recent discussions in the many-body context.

Before we specialize to the case of hydrodynamics, a word on filter functions $f$. One common choice of $f$ is $f(E)=e^{-\beta E}$. It is a matter of Fourier analysis to show that this would give a ramp looking like $\int_{E_{\min}}^{E_{\max}} dE e^{-2\beta E}\frac{T}{2\pi}$. This integral would be dominated by the lowest-energy region of the spectrum. Importantly, the region of the spectrum which in thermodynamics corresponds to inverse temperature $\beta$ is often very heavily suppressed. The physical reason why the coefficient of the ramp doesn't depend on the density of states (and hence why the densest region of the spectrum cannot dominate) can be seen from the equation \eqref{eq:JPDF}. The distribution of the eigenvalues can be interpreted as a Boltzmann distribution where the eigenvalues are particles in some confining potential $NV(x)$ ($V = \frac{x^2}{2}$ for a Gaussian random matrix) that repel each other with potential $2\log|\lambda_i-\lambda_j|$. If we approximate $\rho$ as a continuous density instead of a train of $\delta$ functions, then there is some saddle point density which can be solved from an integro-differential equation involving $V$. The connected SFF, however can be seen as the fluctations in $\rho$. At short times and large energies, where this approximation is valid, we see that the energy function we are trying to minimize is purely quadratic in the $\rho$s, so fluctuations shouldn't depend on the background $\rho$. A longer discussion on this point can be found in \cite{haake2010quantum}.

Because canonical SFFs select only dynamics near the ground state, we need a different approach. If we are interested in spectral statistics far from the ground state, we must instead choose an $f$ like $f(E)=e^{-(E-E_0)^2/4\sigma^2}$, which samples around the energy window of interest. This well-known but unusual property puts spectral statistics in contrast with thermodynamics, where canonical and microcanonical ensembles are equivalent.

In this paper, we discuss the connected SFF. We restrict ourselves to timescales long enough for hydrodynamic effects to kick in. We pay attention only to times much less than the many-body Heisenberg time, after which generic systems have no interesting spectral properties. Our results are non-trivial only at times less than the decay-time for the slowest sound mode, which is quadratic in system length. For a more careful discussion of different timescales in this work, see the end of section \ref{sec:freeStadium}.

\section{Overview of the Hydrodynamic Spectral Form Factor}
\label{section:DPT}

Our approach to the problem is to formulate an effective theory of the SFF contour. We do this by comparing two different but related sets of contours and arguing for a relationship between their effective descriptions. We will suppress the time $T$ and filter function $f$ in many of the expressions below.

\begin{figure}
    \centering
    \includegraphics[scale=0.12]{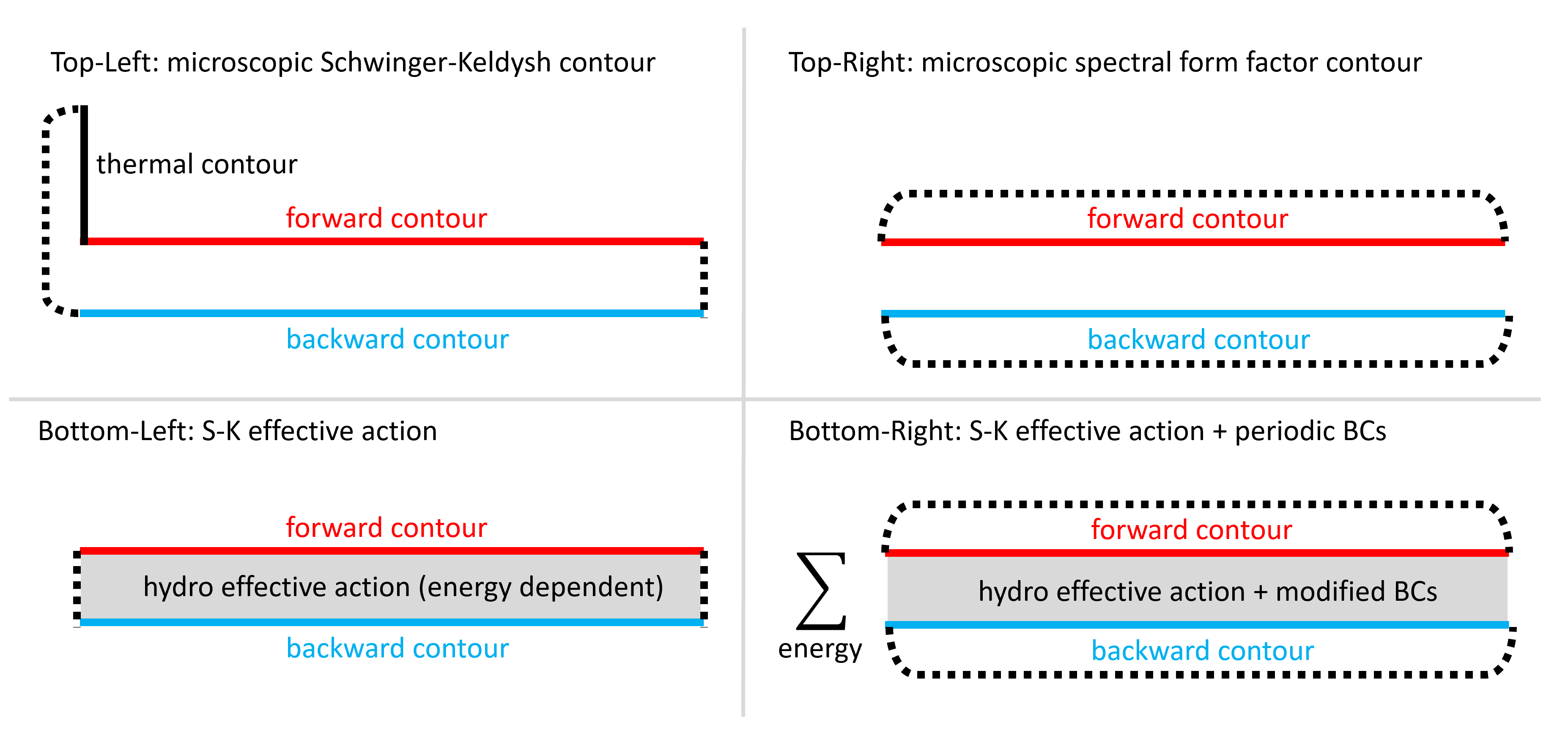}
    \caption{The Schwinger-Keldysh (S-K) contour (left) has a thermal circle of inverse temperature $\beta$ and forward and backwards time evolution legs. By integrating out local microscopic degrees of freedom (bottom), one arrives at an effective theory at the same temperature. The integration couples the left and right legs of the contour. The SFF contour (right) consists of two totally disconnected legs each with periodic boundary condition. By integrating out local microscopic degrees of freedom (bottom), one arrives at the same effective theory as in the Schwinger-Keldysh case (at least up to loop corrections). Remarkably, this integration results in a coupling between the two otherwise decoupled legs. This figure first appeared in \cite{WinerHydro}}
    \label{fig:masterpiece}
\end{figure}

Case 1: The Schwinger-Keldysh (S-K or Kel) contour (Figure~\ref{fig:masterpiece}, left), which computes
\begin{equation}
    Z_{\text{Kel}} = \text{tr}( U[A_1] \rho U[A_2]^\dagger),
\end{equation}
where $\rho$ is the initial state and the $A_i$ are backgrounds fields along the forward ($A_1$) and backward ($A_2$) legs of the contour. In this case, we have a single trace and an initial state which sets the background value of the energy and other conserved charges. 

Case 2: The form factor contour (Figure~\ref{fig:masterpiece}, right), which computes 
\begin{equation}
    \text{SFF} = \tr(f(H) U[A_1]) \tr(f(H) U[A_2]^\dagger),
\end{equation}
where again we allow for different background fields on the forward and backward contours. In this case, we have two traces and no initial state, although we can consider the addition of a filter function $f$ which can be used to select part of the energy spectrum of interest. 

It is important to note that the SFF will be an erratic function of time; we must include some implicit averaging, for example, over time or over an ensemble of Hamiltonians, to give a smooth function of time. We can also compute the same ensemble average of the Schwinger-Keldysh contour, but we will here assume that the SK contour computes self-averaging observables that are only weakly affected by the ensemble average. Henceforth, we compare ensemble averaged versions of Case 1 and Case 2.

The Schwinger-Keldysh contour (Case 1) computes a generating function that gives access, via derivatives with respect to the background fields, to correlators and response functions. It is typically used as a tool to compute dynamical properties out of equilibrium. For our purposes here, we have in mind the Schwinger-Keldysh contour as a tool to compute the long wavelength hydrodynamic response of the system. As we review below, one can formulate an effective theory using the degrees of freedom on the SK contour which can be used to describe the hydrodynamic response of the system.

The SFF contour (Case 2) computes a generalization of the spectral form factor, with the usual form factor being recovered when the background fields are set to zero on each contour. At late times, our expectation is that the SFF will approach that of an appropriate random matrix, but at early times, especially in a system with slow hydrodynamic modes, one can expect systematic deviations from the random matrix result. We are interested in formulating an effective theory for the SFF contour, analogous to the hydrodynamic effective theory of the SK contour.

In a prior work, we gave a proposal for an effective theory of the SFF contour. We now review that proposal and the arguments in favor of it. The basic observation is that the Schwinger-Keldysh and SFF contours can both be viewed as different ways summing matrix elements of the same unitary operator, $U[A_1] \otimes U[A_2]^*$, defined on two copies of the system. For the Schwinger-Keldysh contour (Case 1), we connect the two copies to each other via a link at the initial time which encodes the initial state and a link at the final time which encodes the single trace. For the SFF contour (Case 2), we connect each copy to itself via a link from the initial time to the final time which encodes the two traces.

Based on this clue, we proposed that an effective theory of the Schwinger-Keldysh contour should also give an effective theory of the SFF contour provided the boundary conditions are modified appropriately. As a reminder, this proposal should be viewed as a statement about the ensemble averaged Schwinger-Keldysh contour and the ensemble averaged SFF contour. For the Schwinger-Keldysh contour, the boundary conditions set the value of the total energy (initial state) and introduce a fine-grained correlation between the final states of the two legs (final trace). For the SFF contour, there is no built in correlation between the fine-grained states of the two legs, but the ensemble average produces such a correlation. The SFF contour also does not have a set value of the total energy (or other conserved quantities), so we must integrate over them.

The final proposal is then as follows. Suppose, for a given total energy $E_0$, we have some effective fields $\phi_{1,2}$ which compute $Z_{\text{Kel}}$ as
\begin{equation}
    Z_{\text{Kel}}[A] = \int \mathcal{D}\phi e^{i \int dt d^d x L_{\text{Kel}}[\phi,A;E]},
\end{equation}
where the effective action $W_{SK}$ depends on the background energy $E_0$. Then, we can write a similar effective theory to compute the $\text{SFF}$,
\begin{equation}
    \text{SFF} = \int dE_0 \int  \mathcal{D}' \phi e^{i \int dt d^d x L_{\text{SFF}}[\phi,A;E_0]},
\end{equation}
where 
\begin{itemize}
    \item we now integrate over the total energy $E_0$ (and other conserved charges),
    \item we modify the boundary conditions of $\phi$ to be periodic in time on each contour, 
    \item and we have $L_{\text{SFF}} \approx L_{\text{Kel}}$.
\end{itemize}
The argument for this final point is that the fast decaying degrees of freedom which have been integrated out to yield $L_{\text{Kel}}$ are not significantly affected by the change in boundary conditions provided the time duration $T$ is much larger than the lifetime $\tau_{\text{fast}}$ of any integrated out mode. Performing the same integrating out procedure on the SFF contour thus gives 
\begin{equation}
    L_{\text{SFF}} = L_{\text{Kel}} + \mathcal{O}(e^{-T/\tau_{\text{fast}}}).
\end{equation}

In subsection \ref{subsec:CTP} we quickly review the closed time path (CTP) formalism which deals with the Schwinger-Keldysh contour. In subsection \ref{subsec:DPT} we discuss the doubled periodic time (DPT) effective theory, a cousin of CTP which deals with the SFF contour. Finally in subsection \ref{subsec:soundPole} we discuss a special modification of the CTP and DPT for systems with oscillatory modes.

\subsection{Review of Closed Time Path Formalism}
\label{subsec:CTP}

Hydrodynamics is the program of creating effective field theories (EFTs) for systems based on the principle that long-time and long-range physics is driven primarily by conservation laws and other protected slow modes. One particularly useful formulation is the CTP formalism explained concisely in ~\cite{glorioso2018lectures} and in more detail in ~\cite{crossley2017effective,Glorioso_2017,gao2018ghostbusters}. Other approaches to fluctuating hydrodyamics can be found in~\cite{Grozdanov_2015,Kovtun_2012,Dubovsky_2012,Endlich_2013}.

The CTP formalism lives on the Schwinger-Keldysh contour, pictured in figure \ref{fig:masterpiece}. Its central object of study is the partition function
\begin{equation}
    Z_{\text{Kel}}[A^\mu_1(t,x),A^\mu_2(t,x)]=\tr \left( e^{-\beta H} \mathcal{P} e^{i\int dt d^d x A^\mu_1j_{1\mu}} \mathcal{P} e^{-i\int dt d^d x A^\mu_2 j_{2\mu}}\right),
\end{equation}
where $\mathcal P$ is a path ordering on the Schwinger-Keldish contour. The $j$ operators are local conserved currents. 

For $A_1=A_2=0$, $Z_{\text{Kel}}$ reduces to thermal partition function at inverse temperature $\beta$. Differentiating $Z_{\text{Kel}}$ with respect to the $A$s generates insertions of the conserved current density $j_\mu$ along either leg of the Schwinger-Keldysh contour. Thus $Z_{\text{Kel}}$ is the generating function of all possible contour-ordered correlation functions of current operators. In particular, for systems with a conserved energy, the energy density operator can always be extracted from the hydrodynamic action.

One can write $Z_{\text{Kel}}$ as 
\begin{equation}
      Z_{\text{Kel}}[A^\mu_1,A^\mu_2]=\int \mathcal D \phi^i_1\mathcal D \phi^i_2 \exp\left(i\int dt d^d x L_{\text{Kel}}[A_{1\mu},A_{2\mu},\phi^i_1,\phi^i_2]\right),
\end{equation}
for some collection of local fields $\phi$s. The fundamental insight of hydrodynamics is that at long times and distances, any massive $\phi$s can be integrated out. All that's left over is one $\phi$ per contour to enforce the conservation law $\partial^\mu j_{i\mu}=0$. Our partition function can be written 
\begin{equation}
    \begin{split}
         Z_{\text{Kel}}[A^\mu_1,A^\mu_2]=\int \mathcal D \phi_1\mathcal D \phi_2 \exp\left(i\int dt d^dx L_{\text{Kel}}[B_{1\mu},B_{2\mu}]\right),\\
         B_{i\mu}(t,x)=\partial_\mu \phi_i(t,x)+A_{i\mu}(t,x).
    \end{split}
    \label{eq:AbelianB}
\end{equation}
Insertions of the currents are obtained by differentiating $Z_{\text{Kel}}$ with respect to the background gauge fields $A_{i\mu}$. A single such functional derivative gives a single insertion of the current, and so one presentation of current conservation is the identity $\partial_\mu \frac{\delta Z_{\text{Kel}}}{\delta A_{i\mu}} = 0$.

The effective Lagrangian $L_{\text{Kel}}$ satisfies a number of constraints. The most important is locality. There are no slow modes besides $\phi$, and at long enough distance and time scales integrating out fast modes should yield a local Lagrangian depending on $B_{1,2}$ and their derivatives. There are several additional constraints following from unitarity. They are best expressed in terms of
\begin{equation}
\begin{split}
    B_a=B_1-B_2,\\
    B_r=\frac{B_1+B_2}{2}.
\end{split}
\end{equation}
The key constraints, which will not be proven here but are proven in, say, \cite{crossley2017effective}, are:
\begin{itemize}
    \item All terms in $L_{\text{Kel}}$ have at least one factor of $B_a$, that is $L_{\text{Kel}}=0$ when $B_a=0$.
    \item Terms odd (even) in $B_a$ make a real (imaginary) contribution to the action.
    \item All imaginary contributions to the action are positive imaginary (or zero).
    \item Any correlator in which the chronologically last variable has $a$-type will evaluate to 0 (known as the last time theorem or LTT).
    \item A KMS constraint imposing fluctuation-dissipation relations.
    \item Unless the symmetry is spontaneously broken, all factors of $B_r$ have at least one time derivative. This condition will be lifted in this paper, as we will be considering systems with a spontaneously broken symmetry.
\end{itemize}
For many applications, including calculating SFFs, one typically sets the external sources $A$ to zero, so the action can be written purely in terms of the derivatives of the $\phi$s.

The $\phi$s often have a physical interpretation depending on the precise symmetry in question. In the case of time translation, the $\phi$s are the physical time corresponding to a given fluid time (and are often denoted $\sigma$). In the case of a U(1) internal symmetry, they are local fluid phases. One simple quadratic action which is consistent with the above rules and which describes an energy-conserving system exhibiting diffusive energy transport is (with $\phi_{a,r} \rightarrow \sigma_{a,r}$)
\begin{equation}
    L_{\text{Kel}}^{\text{(diffusion)}}=\sigma_a\left(\kappa \beta^{-1}\partial_t^2\sigma_r-D\kappa \beta^{-1}\nabla^2\partial_t \sigma_r\right)+i\beta^{-2}\kappa(\nabla \sigma_a)^2. 
    \label{eq:lhydro1}
\end{equation}
Here $\kappa$ and $D$ can all be viewed as functions of the background energy $E_0$ which is set by the temperature $\beta$. Later, it will be convenient to also view the $\beta$ appearing in \eqref{eq:lhydro1} as a function of the background energy $E_0$.

\subsection{Review of the Doubled Periodic Time Formalism}
\label{subsec:DPT}
We now explain in more detail how our DPT theory~\cite{WinerHydro}, which is built from the CTP formalism, can be used to compute SFFs. We focus for concreteness on the diffusive action \eqref{eq:lhydro1} as an example. The hydro approach to the SFF predicts that the SFF can be obtained by evaluating the following path integral (still with $\phi_{a,r} \rightarrow \sigma_{a,r}$),
\begin{equation}
    \textrm{SFF}(T,f) = \int_{\mathcal C} \mathcal{D} \sigma_1 \mathcal{D} \sigma_2 f(E_1)f(E_2)e^{i \int dt d^d x L_{\text{Kel}}(\sigma_1,\sigma_2)}.
    \label{eq:SFFhydro}
\end{equation}
Here $\mathcal C$ represents the SFF contour seen on the right of figure \ref{fig:masterpiece}. This contour has two disconnected legs each with real-time periodicity $T$. $\sigma_{1,2}$ represent time reparameterization modes on the two legs of the contour.

The contour in equation \eqref{eq:SFFhydro} should be contrasted with the Schwinger-Keldysh contour. These contours both have two long legs, but very different boundary conditions. The fact that an action initially written to calculate two-point functions in fluctuating hydrodynamics on the SK contour can---when evaluated with different boundary conditions---calculate the spectral form factor is the surprising result of \cite{WinerHydro}.

The key change is in the boundary conditions of the fields. To see this, we first define
\begin{equation}
\begin{split}
    \rho=\frac {\partial L_{\text{Kel}}}{\partial(\partial_t \sigma_a)} = - \kappa \beta^{-1} \partial_t \sigma_r ,
\end{split}
\end{equation}
which can be thought of as the average energy density on the two contours. Consider the spatial- and time-zero-modes of $\rho$ and $\sigma_a$, $\rho_0$ and $\sigma_{a,0}$. $\rho_{0}$ is nothing but the total energy $E_0$ while $\sigma_{a,0}$ is the total relative time shift between the two contours. In the SK contour, $\rho_0$ is set by the initial state, which gives a strongly peaked probability distribution for $\rho_0$. Similarly, in the SK contour, $\sigma_{a,0}$ is fixed to be zero since the times on the two contours are fixed to agree in the far future. By contrast, on the SFF contour the overall energy is not constrained by an initial state. The filter functions can select an energy, and they contribute to the path integral as $f(\rho_0)^2$. (Contributions in which the energies on the two contours are substantially different are suppressed.) Similarly, on the SFF contour the overall time shift is not fixed to be zero, so $\sigma_{a,0}$ should now be integrated over. The domain of integration is periodic since time is identified. In this way, the zero modes produce the expected random matrix ramp,
\begin{equation}
    \text{SFF} \sim \int d\sigma_{a,0} \int dE_0 = T \int d E_0.
\end{equation}

Now we review the calculation in more detail, including both zero- and non-zero-modes. Using the definition of $\rho$, we can rewrite equation \eqref{eq:lhydro1}
\begin{equation}
   L_{\text{Kel}}^{\text{(diffusion)}}=-\sigma_a\left(\partial_t\rho-D\nabla^2\rho\right)+i\beta^{-2}\kappa(\nabla \sigma_a)^2. 
    \label{eq:lhydro2}
\end{equation}
Since the action is entirely Gaussian, we can evaluate the path integral exactly. We first break into Fourier modes in the spatial directions. The integral becomes
\begin{equation}
    \prod_{k}\int\mathcal D\epsilon_k\mathcal D\sigma_{ak}f(E_1)f(E_2) \exp\left(-i\int dt \sigma_{ak}
    \partial_t\rho_k+Dk^2\sigma_{ak}\rho_k-\beta^{-2} \kappa k^2 \sigma_{ak}^2\right) 
\end{equation}
For $k\neq 0$, breaking the path integral into time modes gives an infinite product which evalutes~\cite{WinerHydro} to $\frac{1}{1-e^{-Dk^2T}}$ (and there is indeed no explicit $\kappa$ dependence). For $k=0$, we just integrate over the full manifold of possible $\sigma_a$s and $\rho$s to get $\frac T{2\pi} \int f^2(E_0) dE_0$, up to an overall factor that depends on the measure. So the full connected SFF for systems with a single diffusive mode is
\begin{equation}
   \textrm{SFF}= \frac T{2\pi} \int f^2(E_0) dE_0 \prod_k \frac{1}{1-e^{-D(E) k^2T}} .
   \label{eq:SFFProduct}
\end{equation}
We emphasize again that $D(E_0)$ depends on the background energy $E_0$ and we are integrating over $E_0$. 

Let us now suppose we work in the thermodynamic limit and use a Gaussian filter $f$ to select a particular background energy $E_0$ (up to subextensive fluctuations). In this large-volume limit, the product over non-zero modes can be evaluated by taking the log and then Taylor expanding $\log(1-e^{-Dk^2T})$. The result is
\begin{equation}
    \log \frac{\textrm{SFF}}{\text{SFF}_{\text{zero-modes}}} \approx V\left(\frac{1}{4\pi DT}\right)^{d/2} \zeta(1+d/2),
    \label{eq:diffusiveSFF}
\end{equation}
where $\zeta(s)=\sum_{n=1}^\infty \frac{1}{n^s}$ is the famous Riemann zeta function. This approximation breaks down at times near the Thouless time of the system $T=1/(DL^2)$, where $L$ is the characteristic length. At this point the product over modes in equation \eqref{eq:SFFProduct} can be better approximated as just 1, and the ramp takes on the value one would expect from conventional random matrix theory. The form of \eqref{eq:diffusiveSFF} agrees with the expression derived in~\cite{Friedman_2019} in the limit of large local Hilbert space dimension; the hydro theory predicts \eqref{eq:diffusiveSFF} just given the diffusive dynamics and it can be used to show that that the leading perturbative correction due to hydrodynamic interactions (non-quadratic terms in $L_{\text{Kel}}$) is small when the volume is large.

The for a huge array of systems (including many with sound modes), the product $\prod_k \frac{1}{1-e^{-D(E) k^2T}}$ actually has a physical interpretation as the Total Return Probability (TRP)~\cite{WinerHydro}. If one partitions the configuration space into sectors labeled by $i,j$, then one can define $p_{i\to j}(T)$ as the probability that a system starting in sector $i$ is in sector $j$ at time $T$. The total return probability is
\begin{equation}
    \trp(T)=\sum_{i}p_{i\to i}(T).
\end{equation}
Remarkably, if the sectors are small enough that one can't tell where in a sector the system started after time $T$, then this quantity is very resilient to how exactly sectors are chosen. For instance cutting a sector $i$ into $i', i''$ will replace $p_{i\to i}$ with $p_{i'\to i'}+p_{i'\to i'}=p_{i\to i'}+p_{i\to i''}=p_{i \to i}$. If one chooses to have each sector be a single configuration $\psi$ then the TRP can be written 

\begin{equation}
    \trp=\sum_{\psi} \left|\expval{e^{-iHT}}{\psi}\right|^2.
\end{equation}
The TRP has an interpretation as measuring how much a system still remembers after time $T$. If a system has not spread through configuration space, the TRP will be large, whereas if the system has forgotten its initial configuration the TRP will be one. In purely dissipative systems, the TRP will always be greater than one, and the connected SFF will always be larger than the RMT result. But for systems with some oscillatory character then TRP's behavior can be much more complicated.

That completes our review of the DPT formalism in the context of diffusive dynamics. For a general quadratic hydro theory, the frequencies $\omega=iDk^2$ will be replaced by a more general set of modes $\omega_j(k)$. The parameters specifying this dispersion may also depend on the background energy and the background values of other conserved quantities. The general SFF predicted by the DPT formalism is then
\begin{equation}
    \textrm{SFF}= \frac T{2\pi} \int f^2(E_0) dE_0 \prod_{\textrm{Slow Modes }\phi_j}\prod_{k}\frac{1}{1-e^{i\omega_j(k)T}}.
\end{equation}
We again assume for simplicity that $f$ is chosen to select a particular background energy for which the parameters of $\omega_j(k)$ take some particular value. Then in terms of the random matrix form factor,
\begin{equation}
     \textrm{SFF}_{\textrm{GUE}} =  \frac T{2\pi} \int f^2(E) dE,
\end{equation}
the ramp is enhanced by a factor of 
\begin{equation}
    \Zs = \frac{ \textrm{SFF}}{ \textrm{SFF}_{\textrm{GUE}}} = \prod_{\textrm{Slow Modes }\phi_j}\prod_{k}\frac{1}{1-e^{i\omega_j(k)T}}.
    \label{eq:enhancementDiff}
\end{equation}
The rest of this paper will focus on the computation the product in \eqref{eq:enhancementDiff} for various systems with sound poles. We find our results depend in detail on the geometry of the system, in stark contrast with the diffusive result in equation \ref{eq:diffusiveSFF}, which depends on on volume.

\subsection{Doubled Periodic Time Formalism with a Sound Pole}
\label{subsec:soundPole}
We are finally ready to tackle the problem of sound poles SFF hydrodynamics. To do so, we simple need to specify the frequencies $\omega_j$ entering into \eqref{eq:enhancementDiff}. We use a simple model of sound poles described by the hydro Lagrangian
\begin{equation}
    L_{\text{Kel}}^{\text{(sonic)}} =\frac 12 \phi_a\left(\partial_t^2+\frac{2\Gamma}{c^2} \partial_t^3-c^2\partial_\mu^2\right)\phi_r+\frac{2i\Gamma}{\beta c^2}\phi_a\partial_t^2\phi_a +\textrm{higher derivative terms}.
\end{equation}
This sort of Lagrangian might arise in a superfluid, where $\phi$ plays the role of an order parameter. Alternatively this can be taken as a minimal schematic for a system with a more conventional sound pole, such as familiar fluid systems with energy and momentum conservation.

This system has a characteristic length scale $\ell=\Gamma/c$. Physically, this is the scale below which hydrodynamics breaks down, and corresponds roughly to the mean free path of the constituent molecules.

In general this Lagrangian is cubic in frequency, and the equations of motion gives us three solutions for $\omega$. For small $k$ such that $k\ell\ll 1$, these solutions are  $\omega\approx \pm ck+i\Gamma k^2$, and $\omega \approx i\frac{2\Gamma}{c^2}$. This last mode is fast, and can be ignored in the infrared. Note that the solutions for $i\omega$ are either real or come in complex-conjugate pairs, so the enhancement in \eqref{eq:enhancementDiff} is real as it must be.

Ignoring the fast decaying mode, we finally have our formula for the SFF enhancement in the presence of a sound pole,
\begin{equation}
\log \Zs=-\sum_{k, \mathfrak{s}=\pm}\log(1-\exp\{(i\mathfrak{s}ck-\Gamma k^2)T\}).
\label{eq:enhancement}
\end{equation}
where sums over $k$ are always over the numbers $k>0$ such that $k^2$ is an eigenvalue of the Laplacian in the system holding our fluid.

By Taylor expanding the log on the right, we also have the alternate formula
\begin{equation}
\log \Zs=\sum_{j=1}^\infty \sum_{k^2,\mathfrak{s}}\frac{\exp(j(i\mathfrak{s}ck-\Gamma k^2)T)}j.
\label{eq:enhancementExpanded}
\end{equation}
The presence of the complex exponentials in equation \eqref{eq:enhancementExpanded} leads to qualitatively new behavior not seen in purely diffusive hydro. While systems without sound poles see $\Zs$ decay monotonicly as $T$ increases, in sonic systems we see an intricate interplay between the positive and negative terms in \eqref{eq:enhancementExpanded}.

We will now compute the enhancement in a variety of sonic scenarios. Corrections to the formula \eqref{eq:enhancement} arise from higher-derivative terms and non-Gaussian terms not included in $L_{\text{Kel}}^{\text{(sonic)}}$. Corrections can also arise from the finite width of the filter function.

\section{Babylon: The 1D Sound Pole}

\begin{figure}
\centering
    \includegraphics[scale=0.5]{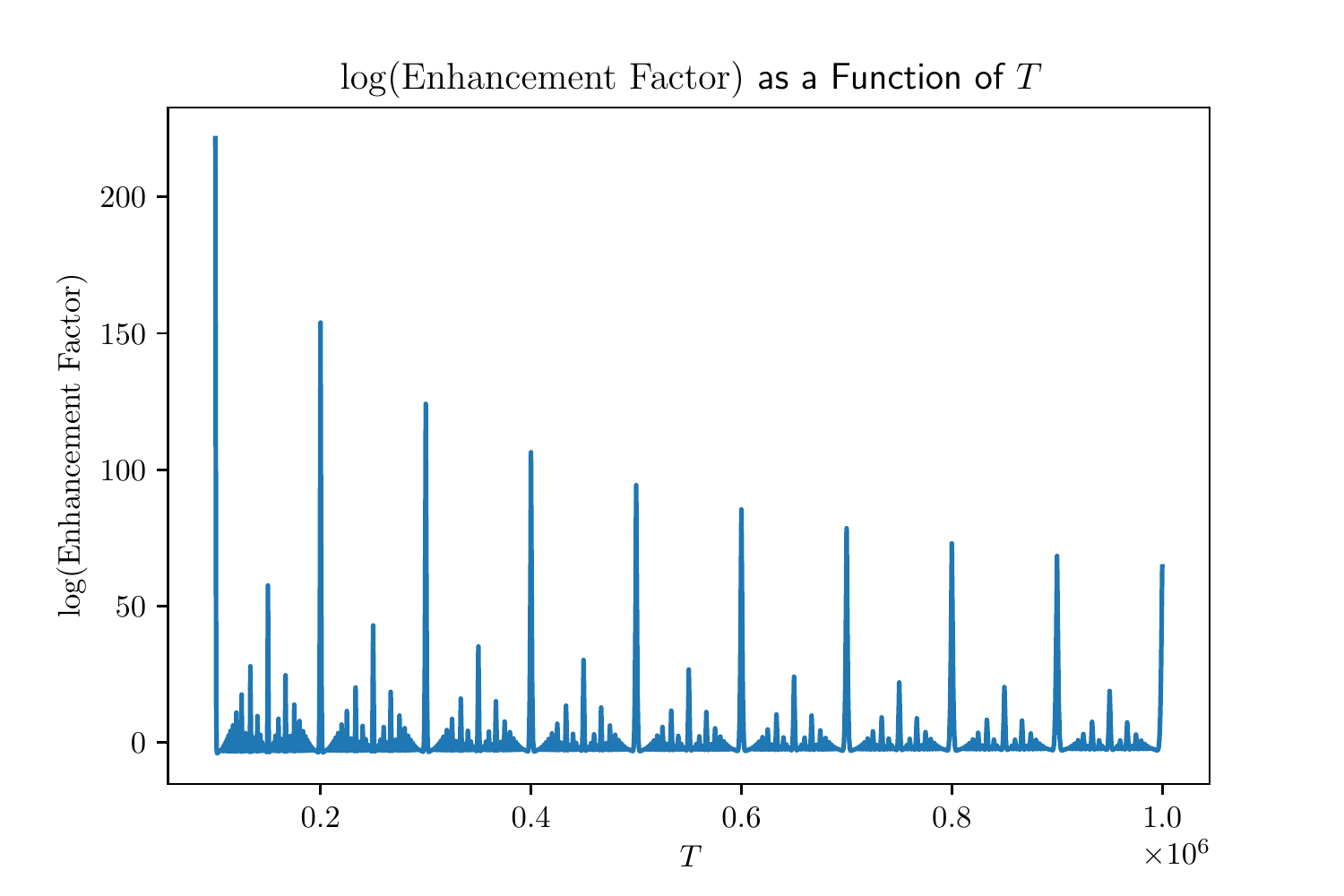}\includegraphics[scale=0.5]{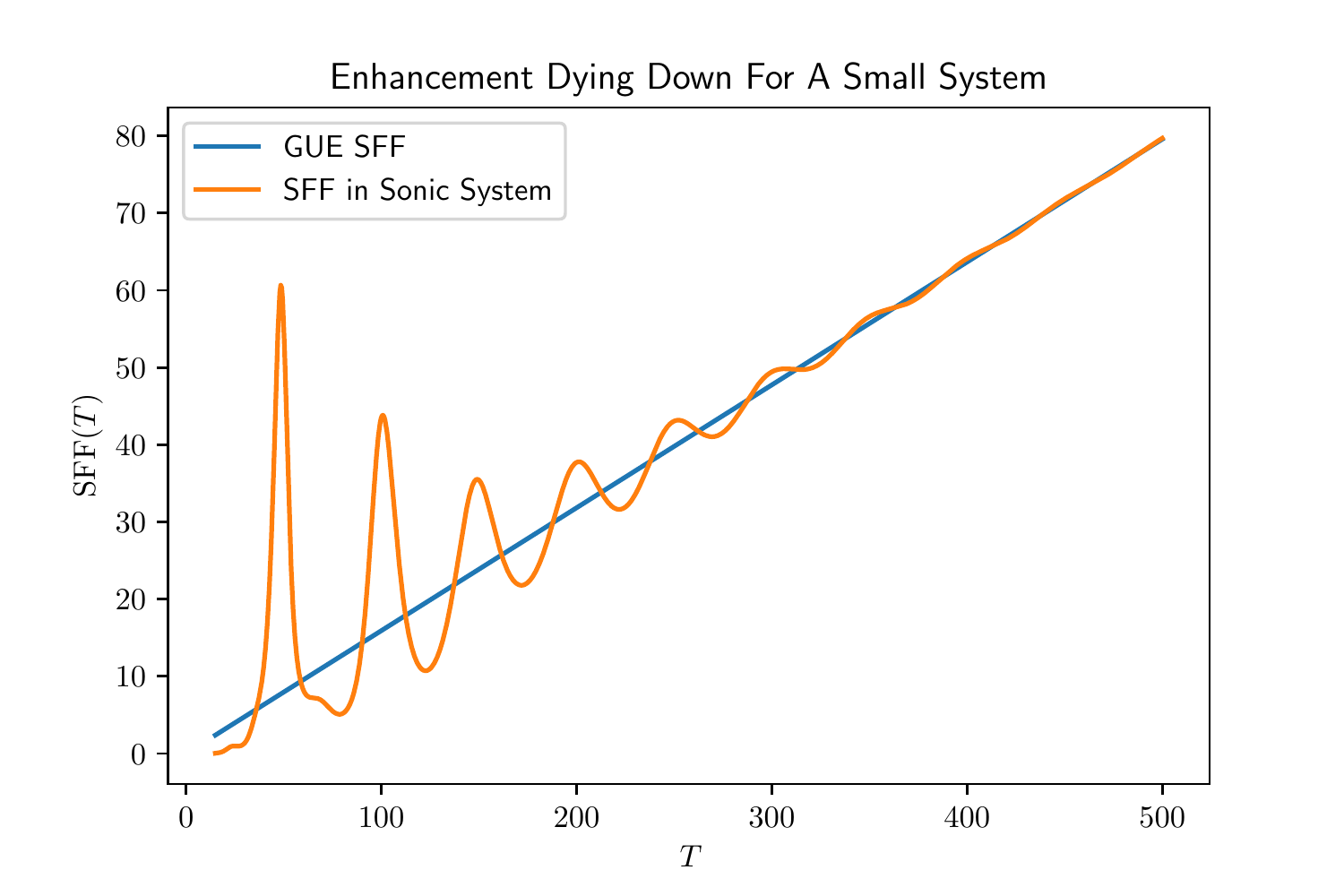}
    \caption{The first image shows the log of the SFF enhancement factor for a system of length $L=10^5$ with $c=1$, $\Gamma=1$, as calculated in equation \ref{eq:SFF1Dproduct}. The graph cuts off at a lower bound of $T=9\times 10^4$, otherwise the $T=0$ divergence would overwhelm the rest of the image. The second graph shows a more sedate choice of $L=50$ $c=1$, $\Gamma=1$. At this smaller scale, only fifty times the mean free path, fewer terms in product \ref{eq:enhancementDiff} and the enhancement is small enough that the sonic SFF and the (connected) GUE SFF can be plotted on the same axes. Note the oscillatory behavior, and the fact that at certain times the sonic SFF is actually smaller than the pure random matrix system.}
    \label{fig:blowup}
\end{figure}

\begin{figure}
    \centering
    \includegraphics{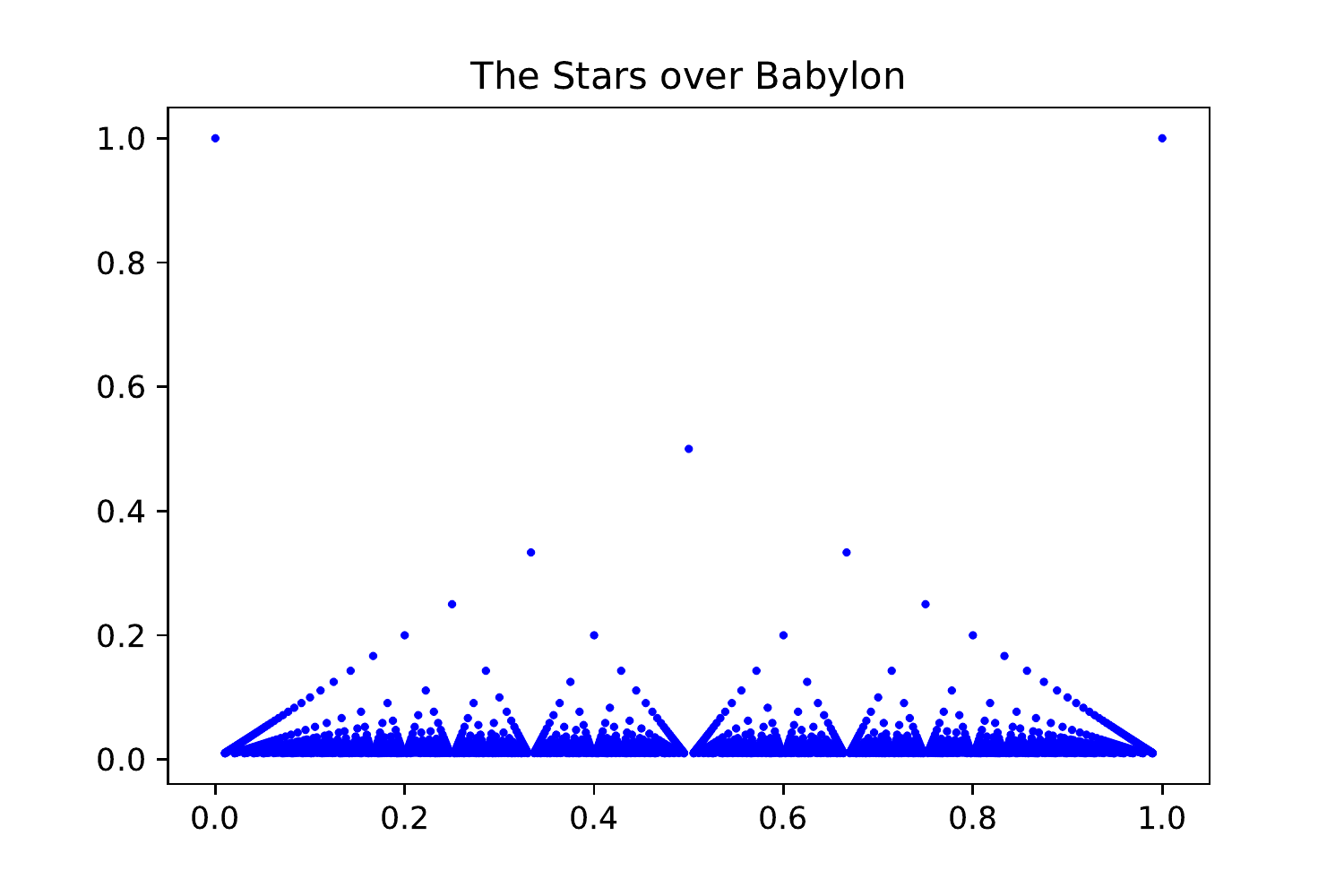}
    \caption{The Stars over Babylon. The Stars over Babylon function in this range corresponds to one period in figure \ref{fig:blowup}.}
    \label{fig:babylon}
\end{figure}

In this section we will concern ourselves with a very specific problem: the SFF enhancement for a 1D system with periodic boundary conditions. In this case, the $k$s are just $2\pi n/L$. In order to get our bearings, let's first evaluate equation \eqref{eq:enhancement} numerically. For a particular parameter choice, the results are shown in figure \ref{fig:blowup}.

Before we can understand this fascinating picture, we need to discuss the function
\begin{equation}
    f(x) = 
\begin{cases}
  \frac{1}{n} &\text{if }x = \tfrac{m}{n}\quad (x \text{ is rational), with } m \in \mathbb Z \text{ and } n \in \mathbb N \text{ coprime}\\
  0           &\text{if }x \text{ is irrational.}
\end{cases}
\label{eq:babylon}
\end{equation}
This function is known by many fanciful names including the popcorn function, the raindrop function, the countable cloud function, and, best of all, the Stars over Babylon.

What does this fairytale function have to do the sums in equations \eqref{eq:enhancement}? The answer is that there is a huge enhancement whenever the system is in resonance, that is whenever $cT/L$ is rational. Setting $k= 2\pi \q/L$ and $cT/L=m/n$ the sum in equation \eqref{eq:enhancement} becomes
\begin{equation}
\begin{split}
    \log \Zs=-2 \sum_{\q>0, \mathfrak{s}=\pm}\log(1-\exp\left\{\left(2\pi i\mathfrak{s} \frac{c\q}{L} -\Gamma \frac{4\pi^2\q^2}{L^2}\right)T\right\})\\
    =-2 \sum_{\q>0, \mathfrak{s}=\pm}\log(1-\exp\left\{\left(2\pi i\mathfrak{s} \frac{m \q}{n} -\Gamma \frac{4\pi^2q^2}{L^2}\right)T\right\}).
    \label{eq:SFF1Dproduct}
\end{split}
\end{equation}
so every $n$th term in the sum over $q$ contributes a term $-\log(1-\exp(-\Gamma k^2 T))$, which is very close to $-\log 0$ when $k$ is small. Since only a $1/n$ fraction of the modes contribute, and contributions like this swamp out any others, we get a structure like equation \eqref{eq:babylon}. A more careful accounting in the next subsection will reveal that the Babylon formula is actually modified to $n^{-3/2}$ from $1/n$. Here $n$, as the denominator in a Babylon-like function, plays the role of a sort of order. Smaller $n$ spikes are more resilient to dissipation while larger $n$ spikes get wiped out more quickly.

\subsection{Details on the 1D Spectral Form Factor}

\begin{figure}
\begin{center}
    \includegraphics[scale=0.8]{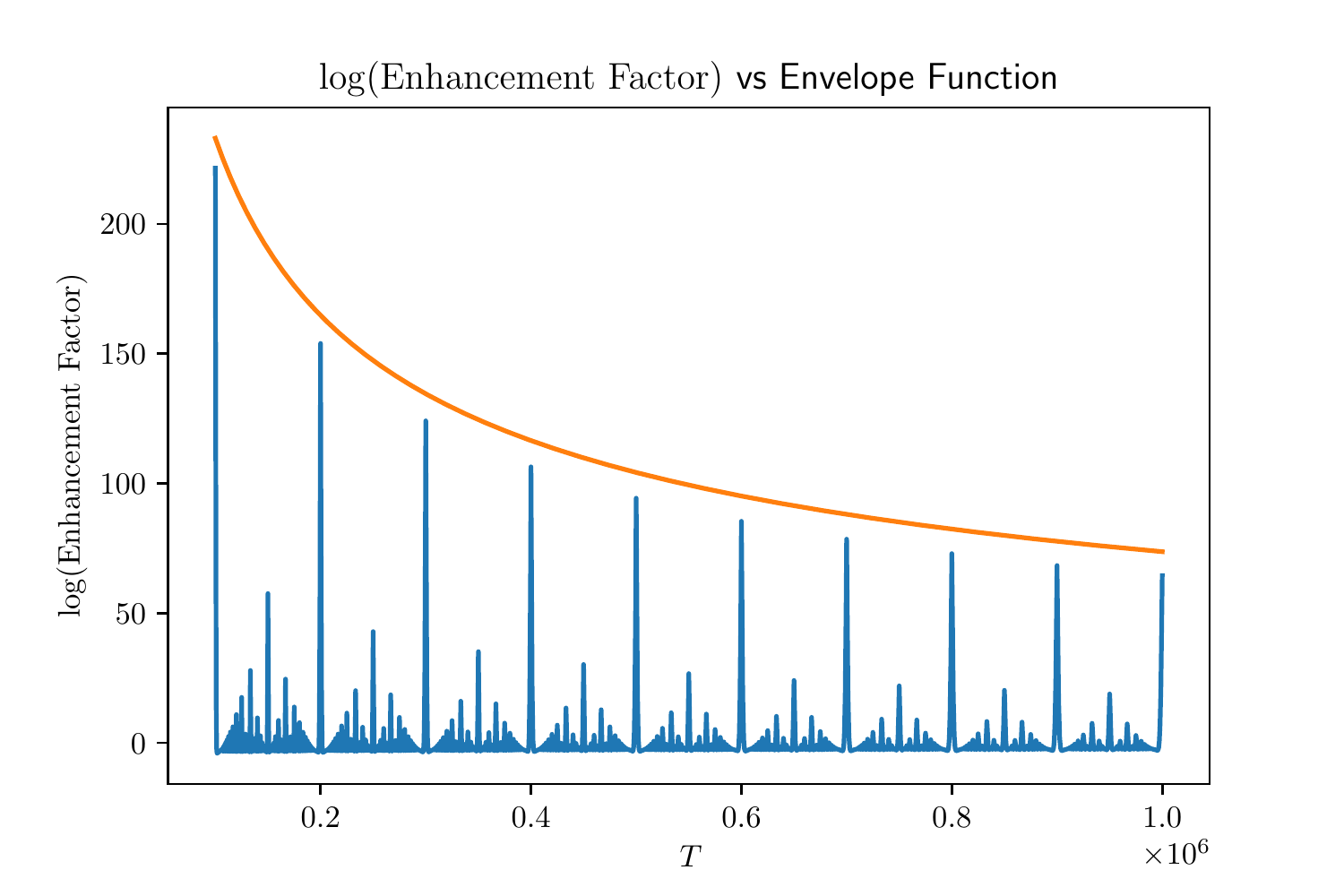}
    \end{center}
    \caption{Equation \eqref{eq:evelope1} does a good job estimating the envelope for $L=1,000,000$, $c=1$, $\Gamma=1$. The source of the discrepancy comes from approximating the sum in equation \eqref{eq:envelopeSum} as an integral. This approximation is valid in the limit $cL\gg \Gamma$ where many terms contribute to the sum.}
    \label{fig:envelope1}
\end{figure}

Not content to observe the Stars over Babylon pattern, let's work out some quantitative details. For instance, how tall should one expect the peaks at full resonance to be? These are the times $T$ satisfying $cT/L=m$ for some integer $m$. So we have
\begin{equation}
    \log \Zs(T=mL/c)=-2\sum_{\q\in \mathbb N/\{0\}}\log(1-\exp(-\Gamma \left(\frac{2\pi \q}{L}\right)^2T))
    \label{eq:envelopeSum}
\end{equation}
If we replace $\Gamma$ with $D$, we see that at these special resonant times, this is also the enhancement factor for normal diffusion. So for times $T\ll L^2/\Gamma$ we can thus use the same methods leading to \eqref{eq:diffusiveSFF} (namely Taylor expanding the log and approximating the sum over $q$s with an integral) to get
\begin{equation}
    \log \Zs \left(T=\frac{mL}c\right)=2 L\left(\frac{1}{4\pi \Gamma T}\right)^{1/2} \zeta(3/2).
    \label{eq:evelope1}
\end{equation}
For a picture of this envelope versus the actual function, see figure \ref{fig:envelope1}.
We can also evaluate an envelope for the shorter peaks for $n>1$. In these cases it is best to break the sum into sets of $n$ consecutive terms. We can write this exactly as 
\begin{equation}
    \log \Zs\left(T=\frac{mL}{nc}\right)=-\sum_{\q\in \mathbb N/\{0\},\mathfrak{s}=\pm}\sum_{\q_2=0}^{n-1}\log(1-\exp(-\Gamma \left(\frac{2\pi (n \q-\q_2)}{L}\right)^2T+\mathfrak s 2\pi i \q_2/n)).
\end{equation}
In the thermodynamic limit $c L \gg \Gamma$ we can treat the Gaussian as constant within each inner sum. We can then use the identity
\begin{equation}
    \sum_{\q=0}^{n-1} \log(1-a\exp(2\pi i\q/n))=\log( 1-a^n)
\end{equation}
to perform the inner sum. We have
\begin{equation}
    \log \Zs\left(T=\frac{mL}{nc}\right)=-\sum_{\q\in \mathbb N/\{0\}}\sum_{\q_2=0}^{n-1}\log(1-\exp(-n\Gamma \left(\frac{2\pi n\q}{L}\right)^2T)).
\end{equation}
This is the same sum again but with $\Gamma$ replaced with $n^3$ $\Gamma$. This means that we have
\begin{equation}
    \log \Zs\left(T=\frac{mL}{nc}\right)=L\frac 1 {n^{3/2}} \left(\frac{1}{4\pi \Gamma T}\right)^{1/2} \zeta(3/2).
    \label{eq:envelopen}
\end{equation}
This $n^{-3/2}$ is interesting and surprising. Again, in the original Babylon function, there is just an $1/n$. We justified this by saying that $1/n$ of the terms contribute exploding positive contributions to the sum. However, the other $n-1$ terms can be shown to on average contribute slightly negative suppressing terms. This provides the intuition for the $n^{-3/2}$ in equation \eqref{eq:envelopen}. 

A final word about when these approximations become valid. Based on numerically evaluating equation \eqref{eq:enhancement}, it seems that the spikes at integers don't become clearly visible until this system size is a thousand relaxation lengths, and the full fractal structure isn't visible until ten thousand. Figures \eqref{fig:blowup} and \eqref{fig:envelope1} use a system size of a million. Needless to say, this is quite beyond the realm of any foreseeable exact diagonalization calculation or experimental technology.

\subsection{A Fourier Perspective and the Width of the Peaks}
\label{subsec:Fourier}
The analysis in the above subsection tells us the height of the peaks, but doesn't tell us anything about their breadth or shape. We can extract this information by taking the Taylor expansion of $\log (1-x)=\sum_{j=1}^\infty -\frac 1j x^j$
\begin{equation}
    \log \Zs=2 \sum_{\q>0, \mathfrak{s}=\pm}\sum_{j=1}^\infty \frac 1j\exp\left(2\pi i\q \mathfrak{s}\frac{cT}{L}j \right) \exp\left(-j\frac{4\pi^2 \Gamma \q^2}{L^2}T\right),
\end{equation}
For each $j$ this is the Fourier series representation of a chain of tight peaks (missing a $\q=0$ term). The peaks have Gaussian shape, and recur after time $\frac L{cj}$. Thus we see that the $j$th term in this sum corresponds only to the peaks at $T=\frac{mL}{jc}$, or in other words we can identify $j$ with $n$ in the previous section.

The total area under the Gaussian from the $j$th sum is given by the amplitude of the missing $\q=0$ term times the period. This is just an amplitude  of $\frac 1j$ times a period of $\frac L {cj}$ for a total area of $\frac{L}{cj^2}$.

Combined with the results of the previous subsection, we know that the width of the Gaussian is something like $\# \frac 1c \sqrt{\frac{\Gamma T}{n}}$. This implies that at sufficiently large $n$, the width of the peaks does become greater than the area between the peaks, which means that the lovely Stars over Babylon structure does not have infinite resolution.

\subsection{Instability to Interactions}
\label{subsec:stardestroyer}
It is also worth investigating how higher derivative terms and interactions would affect the qualitative result of the Babylon-shaped enhancement factor. In the thermodynamic limit, the higher derivative terms don't substantially affect the pattern. The higher derivative terms only affect larger values of $k$, which are already suppressed since they have factors of $e^{-\Gamma k^2 T}$.

Interactions, however, can have a more noticeable impact. A more detailed discussion will need to wait until appendix \ref{sec:Interaction}; we will quote the results of that appendix here. An important point is that while in the CTP formalism the velocity would merely be renormalized by a smooth amount, in DPT the  velocity can be renormalized by an amount depending sensitively on frequency. We hypothesize that this irregularity would substantially derange the intricate pattern, likely in favor of a more erratic pattern of spikes.

If there were some large number of local degrees of freedom, in the tradition of \cite{Chan_2018,Friedman_2019,moudgalya2020spectral,Chan_2022}, then these interactions would be heavily suppressed. But a large local number of degrees of freedom combined with a large system size means a truly large Hilbert space dimensions, making the empirical observation of these patterns a remote possibility for the foreseeable future.

Moreover, in the Schwinger-Keldysh case, 1d sonic hydrodynamics is known to be unstable \cite{spohn_2020}, and flows to the KPZ universality class \cite{KPZ,Krug1997OriginsOS,krajnik_2020}. For SFF hydrodynamics, these concerns are modified. The periodic time changes the significance of diagrammatic corrections in complicated ways \cite{WinerHydro}, and it is unclear in what dimensions weakly coupled hydrodynamics is stable in the IR limit. This will be discussed more in the outlook.

%%%%%%% new %%%%%%%%%%%%%%%%

That said, it is far from obvious that the KPZ scaling completely destroys the Stars. This is because the ballistic propagation of sound, which is responsible for the basic resonance structure, is still present in the KPZ case. Said differently, one can view the flow to the KPZ universality class as modifying the dispersion from $\omega = c k - i \Gamma k^2 + \cdots$ to $\omega = c k - i \tilde{\Gamma} k^{3/2} + \cdots$. This suggests that the KPZ physics most strongly modifies the envelope function and may have a weaker effect on the resonances. We emphasize again that this requires further study and is a non-trivial problem in the DPT formalism, one that combines relevant interactions with periodic time.

However, we can sketch out one starting point for the analysis. One way to interpret $Z_{\text{enh}}$ for the unstable sonic fixed point is via a pair of biased diffusion equations, one for the left movers and one for the right movers. These equations are
\begin{equation}
    \partial_t \rho_\pm = \pm c \partial_x \rho_\pm + \Gamma \partial_x^2 \rho_\pm  + \xi_\pm,
\end{equation}
where $\pm$ refers to the left and right movers and we included a stochastic force $\xi_\pm$ to describe hydrodynamic fluctuations. The $Z_{\text{enh}}$ is obtained as the probability that a given initial condition $\rho_\pm(x,0)$ is recovered at later time at later time $T$. Focusing on the $+$ component, we have
\begin{equation}
    Z_{\text{enh}}(T) = \int D \rho_+(0) D \xi_+ P(\xi_+) \delta[\rho_+^{\text{diff}}(T;\xi_+) - \rho_+(0)],
\end{equation}
where $D\rho_+(0) D\xi$ denotes a functional integral over the noise and the initial condition and $\rho_+^{\text{diff}}(T;\xi)$ solves the biased diffusion equation with noise $\xi$.

The resonance condition, $m L = n cT$, arises as follows. If we ignore the $\Gamma$ term and the stochastic term, then a given profile $\rho_+(x,0)$ is merely translated by the dynamics to $\rho_+(x,T)=\rho_+(x+cT,0)$. Hence, for a localized wavepacket the probability to return is zero unless
\begin{equation}
    m L = c T
\end{equation}
for some integer $m$. The full set of resonances arises by considering perturbations of specific wavelenths, e.g. the $n=2$ case corresponds to perturbations of wavelength $\lambda = L/2, L/4, L/6, \cdots$ that need only be translated by half the length of the system to return to themselves. The effect of the noise and $\Gamma$ term is to broaden these sharp features as discussed above.

Now consider the KPZ case. Because of the peculiarities of 1+1d kinematics, the decomposition into left and right movers still provides an approximate starting point for the analysis. Focusing again on $\rho_+$, the simplest equation which captures the relevant effects is 
\begin{equation}
    \partial_t \rho_+ = c \partial_x \rho_+ + \frac{c'}{2} \partial_x \rho_+^2 + D \partial_x^2 \rho_+ + \xi_+. \label{eq:kpz}
\end{equation}
The new term is the $c'$ term, which turns out to be relevant in the scaling sense. Before proceeding, we emphasize again that it is not clear how the time periodicity modifies the standard analysis and whether the displayed terms are sufficient to capture all the physics of interest in our case. 

With that caveat, the enhancement takes the same form,
\begin{equation}
    \Zs(T) = \int D \rho_+(0) D \xi_+ P(\xi_+) \delta[\rho_+^{\text{KPZ}}(T;\xi_+) - \rho_+(0)],
\end{equation}
where now $\rho_+^{\text{KPZ}}(T;\xi_+)$ solves \eqref{eq:kpz} instead of the biased diffusion equation. We see immediately that the translating effect of the $c \partial_x \rho_+$ term is still present, so a localized wavepacket will still have vanishing return probability unless $m L = c T$. Of course, the other terms are crucial, but as in the diffusive case, their effect is plausibly to broaden these primary resonances rather than to destroy them. On the other hand, the subleading peaks, which arose from profiles with special wavelengths that enjoyed an enhanced translation symmetry, e.g., by $L/2$, are harder to analyze in the non-linear theory given by \eqref{eq:kpz}. Indeed, we cannot analyze the physics wavelength-by-wavelength since the non-linear term couples different wavelengths. We think it is plausible that there could still be enhancements in $\Zs$ corresponding to the subleading resonances, but we cannot say for sure without a more complete analysis of the KPZ return probability.

%%%%%%%%%%%%%%%%%%%%%%

\section{Sound In An Integrable Cavity}
\label{sec:intStadium}
We will now make the jump from quadratic hydrodynamics in 1D to higher dimensions. In this case we are now faced with the choice of what shape our system should take: spherical, toroidal, or something more exotic. While in most cases---including the diffusive hydro SFF---the answers to hydrodynamic questions do not depend on this sort of choice, the sonic hydrodynamic SFF will care about the precise shape we choose.

We see that equation \eqref{eq:enhancement} depends sensitively on the detailed eigenvalues of the Laplacian in our system. In this section we will examine the case where the Laplacian operator in our cavity has Poissonian spectral statistics. This might occur, for instance, if we are doing hydrodynamics in a torus, a rectangular prism, or an ellipsoid. More generally, Poissonian spectral statistics for the Laplacian is thought to describe the case where the cavity, when viewed as a stadium/billiard table, gives rise to integrable dynamics for a particle moving inside the stadium~\cite{berry1977level}. For this reason, we will call this case an integrable cavity. But it is important to remember that while the emergent dynamics of an individual sound mode may be integrable, our hydrodynamic assumption requires that the full many-body dynamics be chaotic in terms of microscopic degrees of freedom.

For a general hydrodynamic system, equation \eqref{eq:enhancement} can be rewritten as 
\begin{equation}
    \Zs=\prod_{k}\frac 1{(1-\exp\{(ick-\Gamma k^2)T\})(1-\exp\{(-ick-\Gamma k^2)T\})}.
    \label{eq:intCoeffFormula}
\end{equation} 
Using the Poissonian statistics, we can calculate the expected value of this enhancement factor. In any Poissonian system, the density of Laplacian ``energy'' eigenvalues in the interval $[E_1,E_1+d E_1]$ is independent of that in region $[E_2,E_2+dE_2]$ for $E_1 \neq E_2$. This independence implies a similar independence for $k\sim\sqrt E$. So we can evaluate the expected value of equation \eqref{eq:intCoeffFormula} by partitioning the spectrum of $k$ into non-overlapping regions $[k_i,k_i+\delta k]$, finding the expected value of the product over eigenvalues in that region, and then multiplying the results together. 

For concreteness, we will treat the values of $k$ as a Poisson process with intensity $\bar \rho(k)$. This means that in any given region $[k,k+dk]$ the product is $[(1-\exp\{(ick-\Gamma k^2)T\})(1-\exp\{(-ick-\Gamma k^2)T\})]^{-1}$ with probability $\bar \rho dk$ (eigenvalue present) and $1$ with probability $1-\bar \rho dk$ (eigenvalue absent). The expected value is thus \begin{equation}
    1+\left(\frac 1{(1-\exp\{(ick-\Gamma k^2)T\})(1-\exp\{(-ick-\Gamma k^2)T\})}-1\right)\bar \rho dk.
\end{equation}
This means that in expectation we can write
\begin{equation}
    \langle{\Zs}\rangle=\exp \left(\int_0^\infty dk \left(\frac 1{(1-\exp\{(ick-\Gamma k^2)T\})(1-\exp\{(-ick-\Gamma k^2)T\})}-1\right)\bar \rho(k)  \right),
    \label{eq:TruePoisson}
\end{equation}
where the angle brackets represent an average over different cavity configurations.

\begin{figure}
    \centering
    \includegraphics{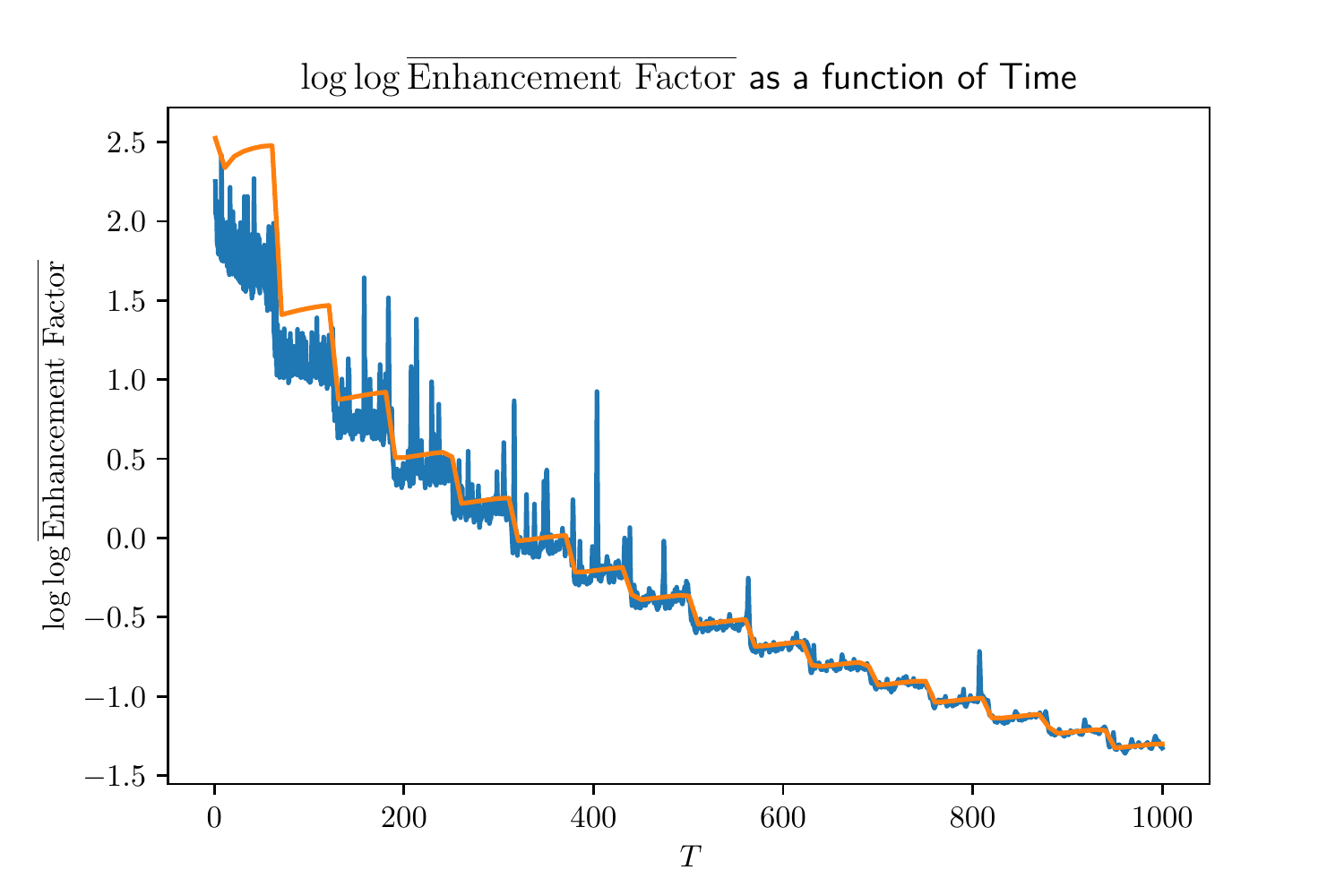}
    \caption{A graph illustrating the agreement of numerical (blue) vs theoretical (orange) predictions for equation \eqref{eq:TruePoisson}. We chose $c=1,\Gamma=0.1$, $\bar \rho(k)=1$. Because equation \eqref{eq:TruePoisson} has a divergence at low $k$ we had a cutoff of $k=0.1$. This cutoff is responsible for the interesting steplike behavior seen in both functions. The blue line is an average of 1971459 samples.}
    \label{fig:TruePoisson}
\end{figure}

This formula isn't the result of an expansion, it assumes only a purely Poissonian density for $k$. Figure \ref{fig:TruePoisson} shows numerics backing up this prediction. To create figure \ref{fig:TruePoisson}, we used the most Poissonian process possible: independent random numbers. It does not correspond to a Laplacian on any particular cavity shape.

It is worth noting that equation \eqref{eq:intCoeffFormula} is a product over many terms. As such, depending on context, the mean value might not be a good representation of typical values, for the same reason that log-normal distributions are not well-clustered around their mean. The expected value of the log of the coefficient can be calculated straightforwardly by taking the log of equation \eqref{eq:intCoeffFormula}.

Another caveat is that real billiard systems, even integrable ones, do not have exactly Poissonian spectral statistics. The clearest exhibition of this is in the case of periodic orbits. To explain how this affects the result, consider a torus of dimensions $2\pi\times 4\pi$. The eigenfunctions of the Laplacian are parameterized by a wavevector of the form $(n_x,n_y)$ where $n_x$ is integer valued and $n_y$ can also take half-integer values. For example, one possible Laplacian eigenstate has wavevector $(1,\frac 12)$, which gives $k^2=\frac 54, k=\frac{\sqrt 5}{2}$. There is another eigenstate with wavevector $(2,1)$, which corresponds to $k=\sqrt 5$. In general, there will be states with $k=n\frac{\sqrt 5}{2}$ for all positive integers $n$. When we multiply the enhancement factors for all of these states, we get the same intricate pattern seen in figure \eqref{fig:blowup}. 

We think that this same effect can exist more generally in any cavity in which a classical particle can take a closed periodic path. To show this fact about the eigenvalues of the Laplacian, we imagine that we start a quantum wavepacket moving under a fictitious Hamiltonian $H_{\text{fict}}=-\frac{1}{2m_{\text{fict}}}\grad^2$ at velocity $v$ around a periodic path of length $L$. The wavepacket does not have a definite energy, but instead a spread of energies well-centered on $E_{\text{fict}}=\frac{k^2}{2m_{\text{fict}}}=\frac{m_{\text{fict}}v^2}{2}$. If we choose a $k$ very large compared to the inverse system size (well into the semiclassical limit) then all of the energy eigenvalues contributing to this wavepacket are close to $E_{\text{fict}}$. 

Because the classical motion is periodic, the wavefunction is going to be approximately periodic in time with period $L/v$.  This means that the wavepacket has overlap with $H_{\text{fict}}$ eigenstates with energies that differ by integer multiples of $\frac{2\pi v}{L}$. So there will be values of $k=\sqrt{2m_{\text{fict}}E_{\text{fict}}}$ spaced out, separated by integer multiples of $\frac{2\pi}{L}$. These evenly spaced out modes violate our assumption of Poissonness, and will lead to contributions like in figure \eqref{fig:blowup}. These effects rely only on the existence of periodic orbits in the cavity, which are are present in both integrable or chaotic cavities.

It is likely, however that these effects will drown out after a comparatively short time once the wavefunction has time to spread out. For chaotic systems this time is known as the Ehrenfest time, and is on the order of $\frac{\log \frac{\textrm{Phase Space Volume}}{h^d}}{\lambda}$, where $\lambda$ is the Lyapunov exponent. For integrable systems the time is given by the same star destroying effects as in subsection \eqref{subsec:stardestroyer}. For a generic integrable system, $\omega$ will depend on $d$ commuting quantum numbers. The dependence will be well approximated as linear, but any higher-derivative terms or interactions in the hydro theory will break that perfect interference.

\section{Sound in a Chaotic Cavity}
\label{sec:freeStadium}
\begin{figure}
    \centering
    \includegraphics[scale=0.8]{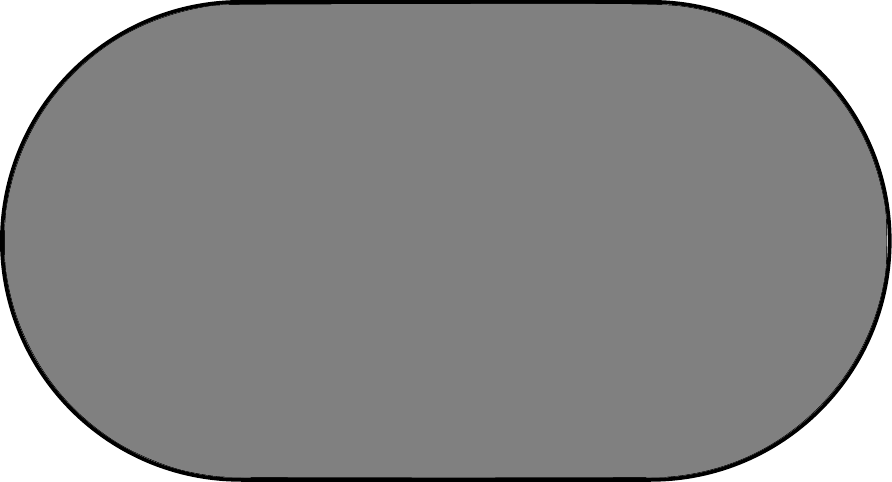}
    \includegraphics[scale=0.8]{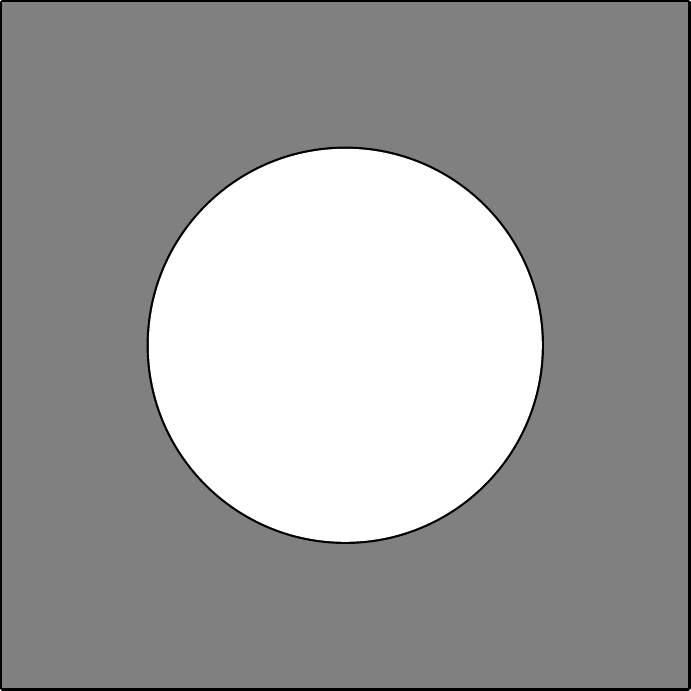}
    \caption{The eigenvalues of the Laplacian are known to have Wigner-Dyson/RMT-like statistics in both Bunimovich Stadium (left) and the Sinai Billiard (right). In this section we explore the spectral statistics of fluids filling up the gray regions in these two shapes.}
    \label{fig:bunSinai}
\end{figure}

In this section we will turn our attention to the case of quadratic hydrodynamic enhancements in a chaotic cavity. As a reminder, the expression for the enhancement from sound poles in a generic cavity is
\begin{equation}
\begin{split}
\log \Zs=-\sum_{k^2, \mathfrak{s}=\pm}\log(1-\exp\{(i\mathfrak{s}ck-\Gamma k^2)T\})\\
=\sum_{j=1}^\infty\sum_{k, \mathfrak{s}=\pm} \frac{1}{j}\exp((i\mathfrak{s}ck-\Gamma k^2)jT)
\end{split}
\label{eq:stadium}
\end{equation}
Let's imagine the shape of the cavity is a Bunimovich stadium or a Sinai billiard (figure \ref{fig:bunSinai}). Then the $k^2$s become the eigenvalues of a level-repelling chaotic Hamiltonian of the Gaussian Orthogonal Ensemble (GOE) universality class. The $k$s are stretched eigenvalues, but still exhibit GOE-type level repulsion.
\begin{figure}
    \centering
    \includegraphics[scale=0.8]{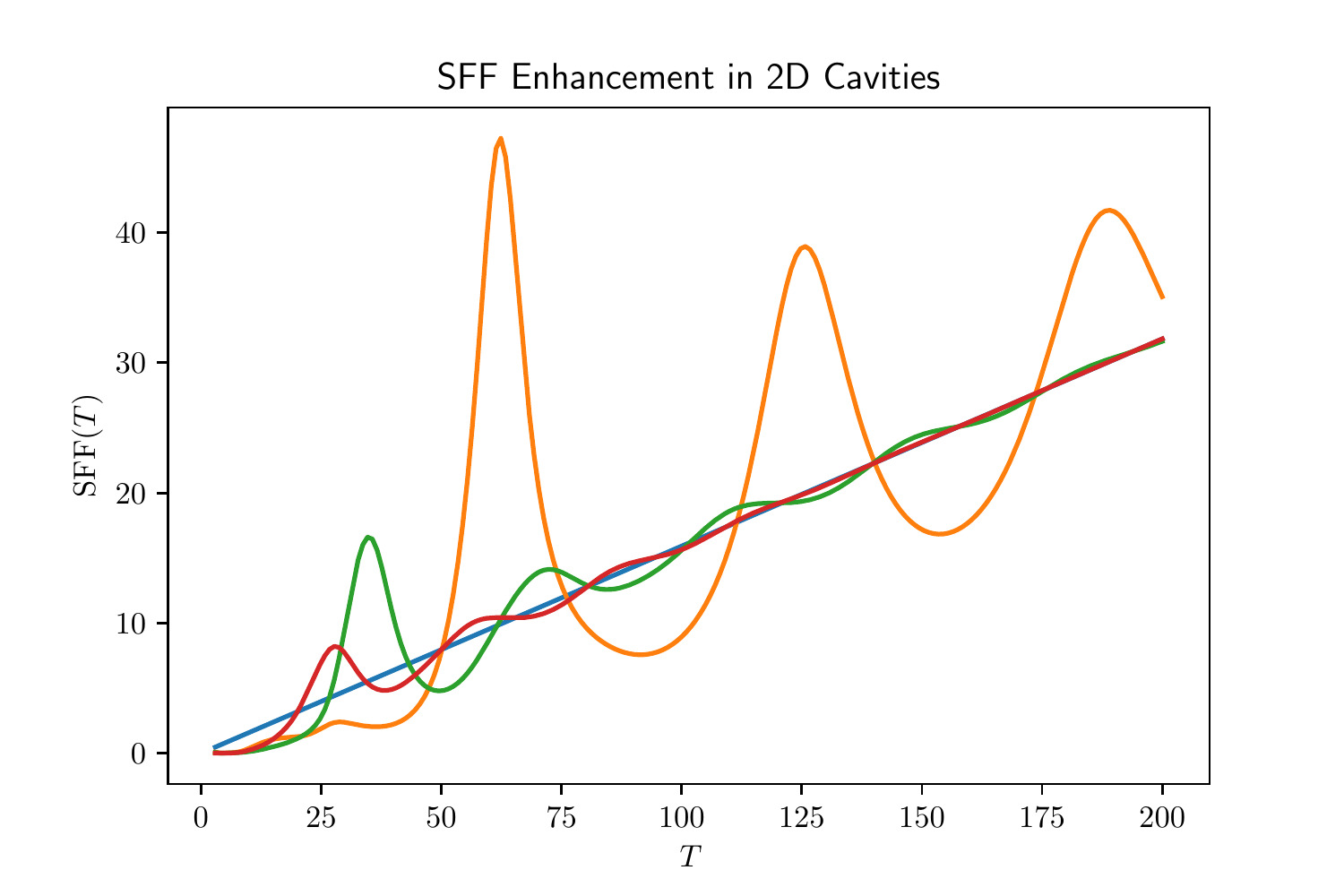}
    \caption{RMT Result (Blue) versus sound-enhanced results in 3 different chaotic cavities with area 40 (Orange, Green, and Red). Notice how the system with one very slow mode (Orange) seems to have fairly different behavior than the faster decaying Green and Reds. As system sizes become larger compared to the mean free path, and more modes can oscillate without dying, interference between various oscillations matters more than the magnitude of any one slow mode.}
    \label{fig:chaosSound}
\end{figure}
We can view the right hand side of equation \eqref{eq:stadium} as a sum over partition functions with imaginary temperature $x$ associated with the $k$ spectrum, 
\begin{equation}
    Z(x,f)=\int dk \rho(k) e^{i k x} f(k)
\end{equation}
where $\rho$ is the exact density of states and $f$ is a filter function. In particular, we have
\begin{equation}
    \log \Zs=\sum_{j=1,\, \mathfrak{s}=\pm}^\infty \frac 1j Z\left[j\mathfrak{s}cT,f_j=\exp(-j\Gamma Tk^2)\right].
    \label{eq:wdef}
\end{equation}
If we fix particular stadium, then we expect that the enhancement factor will be an erratic function of time. To get a smooth result which is calculable, we average over a number of configurations for our stadium (being careful to remain in the GOE universality class) instead of calculating the logarithm of $\Zs$ for a given realization of the stadium. Denoting by $W$ the quantity in equation \eqref{eq:wdef}, we have $\log \langle \Zs \rangle=\log \langle e^W \rangle$, which is by definition the sum of the cumulants of $W$.

Doing a cumulant expansion, we get
\begin{equation}
    \log \langle \Zs \rangle =\sum_\ell \frac 1{\ell!}c_\ell = \mathbb{E}\left[\sum_{j,\mathfrak{s}} \frac 1j Z(j\mathfrak{s}cT,f_j)\right]+\frac 12 \textrm{var}\left[ \sum_{j,\mathfrak{s}} \frac 1j Z(j\mathfrak{s}cT,f_j)\right]+\dots
    \label{eq:basicCumulantExpansion}
\end{equation}
For times smaller than the Heisenberg time of the single-particle system, we can safely approximate the cumulant expansion with just the first two terms, knowing all subsequent terms will suppressed by factors of the cavity volume. We will consider these two terms in turn.

The first cumulant is obtained from the average density of states. Our system has some density of states $\rho(k)$ which fluctuates depending on the precise shape of the cavity. Averaging over cavity shapes we get that $\rho(k)$ fluctuates about some $\bar \rho(k)$. $\bar \rho$ can be calculated in a semiclassical approximation. If the cavity has $d$-dimensional volume $V$, we can expect a density of states
\begin{equation}
    \bar \rho(k)=\frac {V}{(2\pi)^d} S_{d-1}k^{d-1},
\end{equation}
where $S_{d-1}=\frac{2\pi^{\frac{d-1}2}}{\Gamma(\frac {d-1}2)}$ is the surface area of a $d-1$ sphere.

So the expected value of the imaginary-time partition function is
\begin{equation}
    \mathbb{E} \left[\frac 1j Z(j\mathfrak{s}cT,f_j=\exp(-j\Gamma Tk^2))\right]=\frac {V S_{d-1}} {(2\pi)^dj}\int _{0}^\infty dk k^{d-1}\exp(ij\mathfrak{s}cTk)\exp(-j\Gamma Tk^2).
\end{equation}
After including the sum over $\mathfrak{s}$ and relabeling $k \rightarrow -k$ in the $\mathfrak{s}=-$ term, we have
\begin{equation}
     \sum_{\mathfrak{s}} \mathbb{E} \left[\frac 1j Z(j\mathfrak{s}cT,f_j)\right]=\frac {V S_{d-1}} {(2\pi)^dj}\int _{-\infty}^\infty dk |k|^{d-1}\exp(ijcTk)\exp(-j\Gamma Tk^2)
\end{equation}
Right away, we notice that the behavior is qualitatively different for even versus odd $d$. For odd $d$, we have the Fourier transform of an analytic function, while for even $d$, there is a non-analyticity at $k=0$. 

If we evaluate the expression, we get
\begin{equation}
   \sum_{\mathfrak{s}} \mathbb{E} \left[\frac 1j Z(j\mathfrak{s}cT,f_j)\right] =
\begin{cases}
  \frac {V S_{d-1}} {(2\pi)^dj} \sqrt{\frac{2\pi }{j\Gamma T}}(jT)^{1-d}\partial_c^{d-1}\exp(-j\frac{c^2}{4\Gamma}T)&\text{if }d \textrm{ odd},\\
  \frac {2V S_{d-1}} {(2\pi)^dj} \frac{(d-1)!}{(ijcT)^d}+O(T^{-d-1})          &\text{if }d \textrm{ even}.
\end{cases}
\label{eq:dip}
\end{equation}
As we see, in odd dimensions this is an exponential decay, while in even dimensions it is a power-law decay. This is related to the fact that sound waves have sharp edges in odd dimensions and soft edges in even dimensions. The terms in \eqref{eq:dip} can plugged into \eqref{eq:basicCumulantExpansion} and the sum over $j$ computed to get
\begin{equation}
    \log \langle{\Zs}\rangle\supset \begin{cases}
  \frac {V S_{d-1}} {(2\pi)^d} \sqrt{\frac{2\pi }{\Gamma T}}(T)^{1-d}\partial_c^{d-1}\exp(-\frac{c^2}{4\Gamma}T)&\text{if }d \textrm{ odd},\\
  \zeta(d+1)\frac {2V S_{d-1}} {(2\pi)^d} \frac{(d-1)!}{(icT)^d}+O(T^{-d-1})          &\text{if }d \textrm{ even}.
\end{cases}
\label{eq:1ParticleDip}
\end{equation}

The odd dimension result is zero at times greater much than $\frac{\Gamma}{c^2}$. Unfortunately, before this time hydrodynamics is dominated by higher derivative corrections, so in odd dimensions equation \ref{eq:1ParticleDip} must be approached with caution. This fast decay of the exponential-in-volume enhancement factor can be contrasted with the diffusive result quoted in equation \ref{eq:diffusiveSFF}, which decays slowly in all dimensions.
In contrast the rapid decay in odd dimensions, the even-dimensional part of equation \ref{eq:1ParticleDip} is large until times extensive in the system length, and could, in principle, be observed. But it still decays far faster than equation \ref{eq:diffusiveSFF}.

If we stopped here, at the first term in the cumulant expansion, we would be making the approximation $\log \langle{\Zs}\rangle=\langle{\log \Zs}\rangle$. By analogy with the glass literature, we refer to $\log \langle{\Zs}\rangle$ as the annealed average and $\langle{\log \Zs}\rangle$ as the quenched average. The above approximation is thus analogous to the approximation that quenched equals annealed. Typically, the quenched value of a partition function is the value for a typical realization of disorder, whereas the unquenched or annealed value is dominated by more extreme terms. If one were to take a small number of cavity shapes and evaluate the SFF enhancement coefficients, most of the coefficients would look like equation \eqref{eq:1ParticleDip}. This is because we have $\log \Zs \sim \rho$ sample by sample, so the quenched average contains only a linear-in-$\rho$ term whereas the annealed average has contributions from higher cumulants. Equation \eqref{eq:1ParticleDip} also gives the quenched answer for an integrable system.

Now onto the second cumulant, which is
\begin{equation}
    \textrm{var}\left[\sum_{j,\mathfrak{s}} \frac 1j Z(j\mathfrak{s}cT,f_j)\right]=\sum_{j,j',\mathfrak{s},\mathfrak{s}'} \frac{1}{j j'}\int dk dk' \mathbb{E}(\rho(k)\rho(k'))_{\text{conn}} e^{i j k \mathfrak{s} c T + i j' k' \mathfrak{s}' c T} f_j(k) f_{j'}(k'),
\end{equation}
where $\mathbb{E}(\rho(k)\rho(k'))_{\text{conn}}$ is the connected pair correlation function. Further progress can be made under the assumption that the connected correlator has only weak dependence on $k+k'$. In this case, only terms with $j=j'$ and $\mathfrak{s}=-\mathfrak{s}'$ contribute, giving
\begin{equation}
    \textrm{var}\left[\sum_{j,\mathfrak{s}} \frac 1j Z(j\mathfrak{s}cT,f_j)\right] \approx \frac{1}{j^2} \sum_{j,\mathfrak{s}} \int dk dk' \mathbb{E}(\rho(k)\rho(k'))_{\text{conn}} e^{i j \mathfrak{s} (k-k')  c T} f_j(k) f_j(k').
\end{equation}

Now, this expression is essentially a sum over single-particle spectral form factors evaluated at the times $j c T$ and with filter function $f_j$. For each of these we may use the standard GOE result (assuming the single particle SFF is always in the ramp phase) to obtain
\begin{equation}
    \textrm{var}\left[\sum_{j,\mathfrak{s}} \frac 1j Z(j\mathfrak{s}cT,f_j)\right] \approx \sum_j \frac{2}{j^2} \int_0^\infty d\bar{k} \left( \frac{j c T}{\pi} \right) \exp( - 2 j \Gamma T \bar{k}^2).
\end{equation}
The integral is straightforward, and the total variance is thus
\begin{equation}
    \textrm{var}\left[\sum_{j,\mathfrak{s}} \frac 1j Z(j\mathfrak{s}cT,f_j)\right] \approx \sum_j  \frac{c T}{j \pi}\sqrt{\frac{\pi}{2 j \Gamma T}} = \zeta(3/2) \sqrt{ \frac{c^2 T}{2 \pi \Gamma}}. \label{eq:varTerm}
\end{equation}
This answer is eerily reminiscent of equation \eqref{eq:evelope1}. But most interestingly, it is a highly universal contribution to the ramp of a hydrodynamic system with a sound pole. It doesn't depend on any details of the shape of the system, or even its overall size. Just that the sound waves propagating around experience chaotic dynamics.

Equation \eqref{eq:varTerm} increases indefinitely, and it is worth knowing when exactly it starts to fail. The answer is that at times on the order of the single-particle Heisenberg time (inverse level spacing of the single-particle system) the second cumulant reaches a plateau, and cumulants after the second stop being negligible. After a few Heisenberg times, there is no longer any residue of the single-particle level repulsion and the enhancement for chaotic billiards should resemble that for integrable billiards. This, in turn, falls to 1 as the various sound modes die down, long before the full many-body Heisenberg time.

In summary, here is a list of important time scales for the connected sound pole SFF in a chaotic cavity:
\begin{itemize}
    \item Hydrodynamic scattering time of order $\Gamma/c^2$. This is the time-scale after which hydrodynamics becomes a valid approximation.
    \item Single particle Thouless time, on the order of $L/c$. This is the time scale at which equation \eqref{eq:varTerm} becomes a good approximation.
    \item The single particle Heisenberg scale on the order of $\left(\frac 1c V \Gamma^{-\frac{d-1}{2}}\right)^{\frac{2}{d+1}}$. Equation \eqref{eq:varTerm} is dominated by modes with $k$ on the order of $k\sim (\Gamma T)^{-1/2}$. The single particle Heisenberg scale is the time at which $T$ exceeds the density of states in this region. Above this time scale a chaotic billiard should behave like an integrable billiard.
    \item The lifetime of the slowest sound modes $L^2/\Gamma$. This can also be thought of as a many-body Thouless time. After this time the ramp looks like the pure random matrix theory result. 
    \item The many-body Heisenberg time at order $e^{S}$. This is the time scale at which the SFF of the full many-body system plateaus.
\end{itemize}

\section{Discussion and Outlook}

In this paper, we expanded the theory of the hydrodynamic spectral form factor to the more realistic case of hydrodynamics with a second-time derivative. This more sophisticated theory and its accompanying sound pole structure allowed a new phenomenon: interference in the emergent hydrodynamic SFF formalism. We discovered a variety of different interference patterns, all connecting intimately with the SFF of a single particle problem in various cavities. We started with the simplest case, the (necessarily integrable) dynamics of a sound mode in a single dimension, and got the remarkable function in figure \ref{fig:blowup} in the absence of dispersion. In the case of higher dimensional cavities, our results mostly concerned disorder-averages over cavities. Using Wigner-Dyson versus Poissonian statistics for the eigenvalues of the Laplacian of the cavity, we derived the results of sections \ref{sec:intStadium} and \ref{sec:freeStadium}. We derive equations \eqref{eq:TruePoisson} and \eqref{eq:basicCumulantExpansion} respectively, and graph the predicted enhancements for toy examples of the laplacian spectral statistics.

%%%% here %%%%%%%%

Throughout this work, we find that the expected spectral form factor depends sensitively on both the dimensionality of the system as well as whether the cavity supports chaotic or integrable billiard dynamics. For example, equation \eqref{eq:1ParticleDip} take on entirely different forms in odd versus even dimensions, for the same reason that sound has a non-analytic shockwave in odd but not even dimensions. We also note that, if momentum is exactly conserved, then there will be distinct blocks in the many-body Hamiltonian labelled by the many-body momentum (similar to the discussion of conserved quantities in \cite{Winer_2022}).

This work opens the gates for calculations in a wide array of settings, including CFTs (which necessarily have both momentum and energy conservation) and systems with spontaneously broken symmetry. In particular, CFTs on spheres have been shown~\cite{Freivogel_2012} to have eternal non-decaying hydrodynamic modes. This follows from the conformal symmetry, where certain ladder operators $L^+$ satisfy the commutation relation $[H,L^+]=\frac 1R L^+$. Just like in a simple harmonic oscillator, this leads to two-point functions which oscillate exactly, forever, even at times much longer than the Heisenberg time of the system.
Each primary operator with energy $\frac \Delta R$ sits at the bottom of a tower of states which contribute $\frac{e^{i\Delta T/R}}{(1-e^{iT/R})^D}$ to the SFF. These $\frac 1{(1-e^{iT/R})^D}$ factors would lead to strong enhancements in the SFF out to arbitrary times, even deep into the plateau region. It would be interesting to look for a hydrodynamic explanation for these effects, and relate them to the above formalism.

As our ability to numerically calculate the SFFs of quantum fields theories improves \cite{luca}, one might look for at least the leading peaks in a numerical spectral form factor. One highly optimistic possibility is that signatures of these results could be found in the original home of Wigner-Dyson statistics: atomic nuclei. These nuclei are hydrodynamic systems with vibrational modes, these modes might have a signature in the spectral form factor.

\begin{figure}
    \centering
    \includegraphics[scale=0.5]{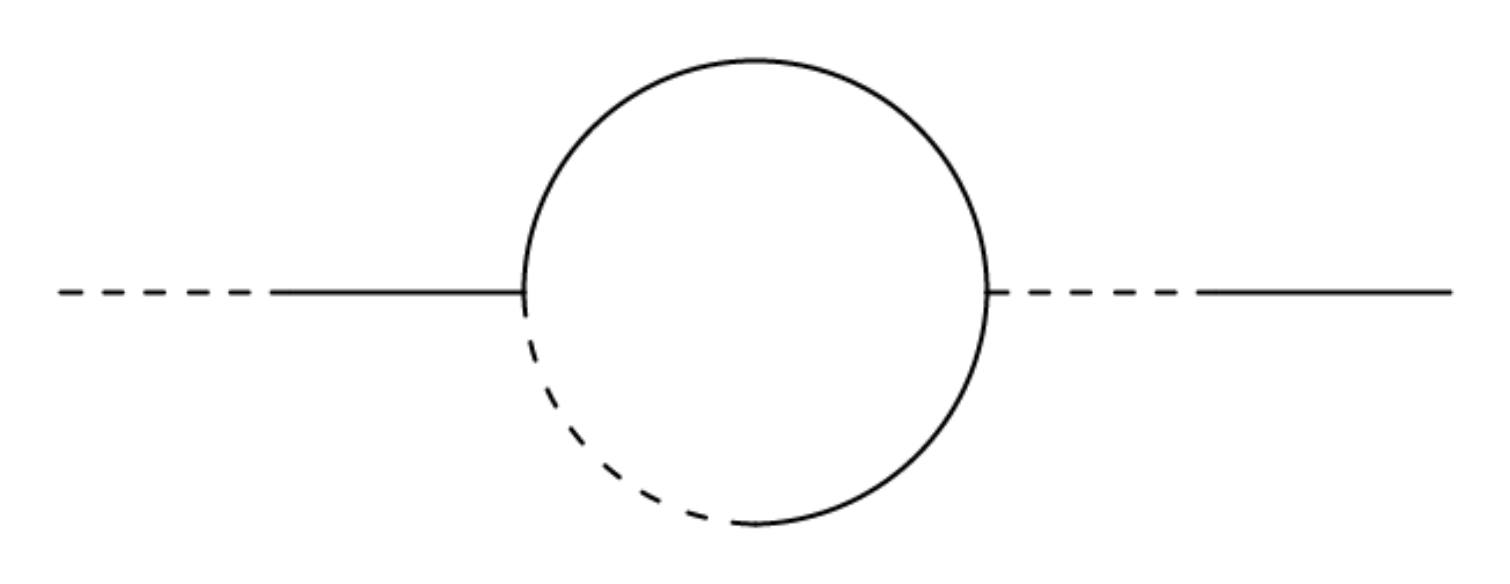}
    \caption{The loop integral in this diagram depends on the second moment of the spectral density.}
    \label{fig:twoPointDiagram}
\end{figure}

An important direction for future work is seeing whether this emergent spectral effect can give rise to enhanced 2-point functions in the CTP hydrodynamic theory. Figure \eqref{fig:twoPointDiagram} shows a diagram which depends on an internal integral over the frequencies of the two legs. If the two-point function of these frequencies is not analytic (as is the case for both integrable and Wigner-Dyson spectral statistics), that leads to a long-time tail (albeit suppressed by a factor of volume).
Finally, the fate of DPT hydro in the long-range limit is still an open question. It is known that in conventional hydrodynamics, the 1D theory is always strongly coupled. In the DPT theory, each leg in each Feynman diagram is modified due to each mode wrapping around the periodic time (see equation \ref{eq:GDPT}). One could imagine these modes destructively interfering at times $T$ incommensurate with the period of the sound modes, thus rescuing the diffusive theory. On the other hand, it is known \cite{WinerHydro} that for purely diffusive dynamics the additional wrapping makes periodic-time hydro strongly coupled even up to two dimensions. We leave this question for future work.

This work was supported by the Joint Quantum Institute (M.W.) and by the AFOSR under FA9550-19-1-0360 (B.G.S.).

\appendix

\section{More On The Total Return Probability}
\label{app:TRP}
\begin{figure}
    \centering
    \includegraphics{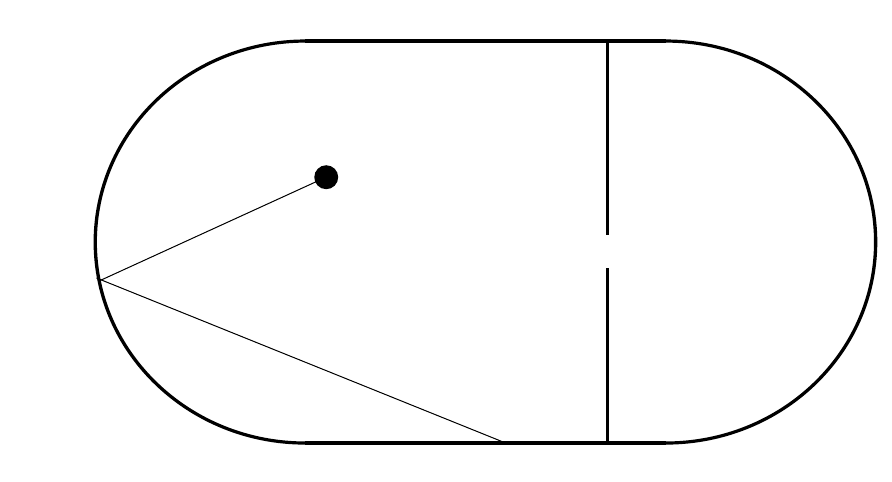}
    \caption{A simple example of chaotic dynamics with two slowly thermalizing sectors. The ball will quickly spread out within the Hilbert Space of a given chamber, but only much more slowly will it transfer between the two sectors. During this time the connected SFF will be larger than the RMT prediction.}
    \label{fig:bunSectors}
\end{figure}
The Total Return Probability (TRP) is a measure of how much a system has thermalized after time $T$. Remarkably, it has also been shown \cite{WinerHydro} to be the enhancement factor for the connected SFF at time $T$. One simple example of a system with slow thermalization is a single quantum particle bouncing around a partitioned Bunimovich stadium in figure \ref{fig:bunSectors} (not to be confused with the much more complicated system discussed in section \ref{sec:freeStadium}, a many-body system filling a Bunimovich stadium). The two chambers of this stadium are our two sectors. This particle will bounce around and quickly (after the Ehrenfest time) be spread out within a given sector. At this point $p_{1 \to 1}$ and $p_{2 \to 2}$ will both be 1 and the TRP will be 2. As more time passes, both $p_{1 \to 1}$ and $p_{2 \to 2}$ will decrease towards the equilibrium probabilities of being in chamber 1 or 2 respectively. Eventually the TRP will be one.
More RMT-focused readers might prefer the example where $H$ is a matrix with
\begin{equation}
    H=\begin{pmatrix}
    H_1&H_{\textrm{small}}\\
    H_{\textrm{small}}^\dagger&H_2
    \end{pmatrix}
\end{equation}
where some weak coupling $H_{\textrm{small}}$ induces jumping between two RMT-like sectors. In this case the TRP can be calculated using Fermi's golden rule. We get
\begin{equation}
    \trp(T)=1+e^{-\lambda T}
    \label{eq:TwoBlockTRP}
\end{equation}
for a specific value of $\lambda$ characterizing the strength of the off-block-diagonal matrix elements.
\begin{figure}
    \centering
    \includegraphics[scale=0.5]{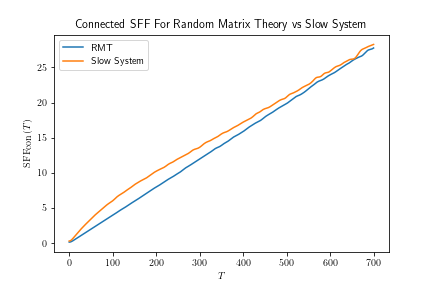}
    \includegraphics[scale=0.5]{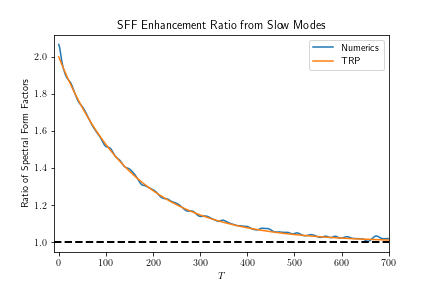}
    \caption{(Left) the RMT vs nearly block diagonal connected SFFs during the ramp regime. (Right) A comparison of the numerical ratio (blue) versus the TRP in equation \ref{eq:TwoBlockTRP} (orange).}
    \label{fig:twoSFFs}
\end{figure}

A more complicated case would be a single particle diffusing through a complicated medium. In this case the sectors would be regions of space and $p_{i \to i}$ would be calculated by solving the diffusion equation. This procedure would agree with previous results obtained through sigma model methods\cite{Kravtsov1994LevelSI,andreev1995spectral}. An even more complicated case would be a many-body system with a conserved charge, where the sectors would be coarse-grained charge distributions $\rho_i(x)$. $p_{i \to i}$ is calculated using fluctating hydrodynamics as discussed in subsection \ref{subsec:DPT} and in more detail in \cite{WinerHydro}. The results agree with a large-$q$ Floquet calculation done in \cite{Friedman_2019}.
\section{SYK2 At Finite Temperature}
\label{app:SYK2}

In \cite{PhysRevLett.125.250601,Winer_2020} the authors investigated the many-body SFF in a specific system of free-fermions with random two-body interactions. By specifically studying the many-body SFF of these
They find an exponential ramp in the infinite-temperature SFF, in particular with a ramp of the form $SFF\propto N^{\#T}$, where the number depended on the specific model under consideration. This exponential ramp behavior can be argued to come from the same cumulant expansion as equation \eqref{eq:varTerm}.

In the case of the SYK2 model with no charge/number conservation, we have
\begin{equation}
    H=\frac 12 \psi_iM^{ij}\psi_j,
\end{equation}
where $M$ is an anti-symmetric real matrix. One can show that the eigenvalues of $M$ follow a semicircle law, and obey GUE level repulsion (the reason for this is that even though the matrix is real, the eigenvalues come in pairs $\pm \lambda$, and there is repulsion for both the positive and negative eigenvalues). We can write the full-system SFF at inverse temperature $\beta$ as
\begin{equation}
    \log SFF=\sum_{\lambda>0}2\textrm{Re}\log(1+\exp(-\beta \lambda+iT \lambda))=\sum_{j\neq 0}\frac{(-1)^j}j\sum_{\lambda>0}\exp(-\beta j \lambda+ijT \lambda))=\sum_{j\neq 0}\frac{(-1)^j}{j}Z(jT,j\beta)  
\end{equation}
Note the similarity between this form and equation \eqref{eq:stadium}. And while at infinite temperature all terms of the cumulant expansion are needed, we will show that at finite temperature only the first two are important. 

For most random matrix ensembles, we can write the probability density as
\begin{equation}
    dP(M)=\exp(-N\tr V(M))dM.
\end{equation}
In terms of the eigenvalues density $\rho(\lambda)=\frac 1N\sum_i \delta(\lambda-\lambda_1)$ and the eigenvalues distribution density as 
\begin{equation}
    dP(\rho)\propto \exp(-\int N^2 V(\lambda)\rho(\lambda) d\lambda+\int N^2\rho(\lambda_1)\rho(\lambda_2)\log|\lambda_1-\lambda_2|d\lambda_1d\lambda_2+\int N\rho(\lambda) \log \rho(\lambda)d\lambda)\mathcal D \rho
    \label{eq:measure}
\end{equation}
At large $N$, this is the quadratic functional
\begin{equation}
    dP(\rho)=\exp(-\int N^2 V(\lambda)\rho(\lambda) d\lambda+\int N^2 \rho(\lambda_1)\rho(\lambda_2)\log|\lambda_1-\lambda_2|d\lambda_1d\lambda_2)\mathcal D \rho.
    \label{eq:measureQuad}
\end{equation}
This means that we can analyze quantities related to $\rho$ by linear transformations (including $Z(T,f)$, or the resolvent $R$) using only the first two cumulants. This breaks down only when $\rho$ gets extremely close to zero. Let us start with $\rho=\bar \rho$, the saddle point, and see what temperatures cause trouble.

For instance, the SFF of the SYK2 model can be written
\begin{equation}
    SFF=\int dP(\rho) \exp(N\int \rho \log(1+\exp(iT\lambda-\beta \lambda))d\lambda+cc)
\end{equation}
We can make use of
\begin{equation}
    \log(1+e^{ix})=\sum_n \log \frac{x-(2n+1)\pi}{(2n+1)\pi}
\end{equation}
This means we can write the SFF as 
\begin{equation}
    SFF=\int dP(\rho) \exp(N\int \rho \sum_n \log \frac{x-(2n+1)\pi/(T+i\beta)}{(2n+1)\pi/(T+i\beta)}+\log \frac{x-(2n+1)\pi/(T-i\beta)}{(2n+1)\pi/(T-i\beta)}d\lambda)
\end{equation}
Combining with the quadratic large $N$ measure in equation \eqref{eq:measureQuad} we can solve for the new saddle-point value of $\rho$. We have
\begin{equation}
\begin{split}
    \rho=\rho_0+\delta \rho\\
    \delta \rho=\frac 1 N \sum_n (2n+1)\frac {\beta}{T^2+\beta^2} \frac{1}{(x-\frac{(2n+1)\pi T}{T^2+\beta^2})^2+(2n+1)^2\pi^2\frac {\beta^2}{(T^2+\beta^2)^2}}
\end{split}
\end{equation}
In the limit of small $\beta$, this is just a train of delta functions of mass $\frac 1 N$. The deepest well will be of depth $\frac {T^2+\beta^2}{N\pi^2 \beta}$. If this becomes $O(1)$, comparable to $\rho_0$, the cumulant expansion breaks down. For any fixed nonzero $\beta$ this isn't a concern, except at very long times. But if $\beta \sim N^{-1}$, we get the more complicated $N^{\#T}$ behavior.

In the regime where the cumulant expansion works, and where $T\gg \beta $, we can evaluate the first two terms $c_1,c_2$ in the cumulant expansion. The first is non-universal and dependent on the density of states:
\begin{equation}
    c_1=\textrm E\sum_{j\neq 0}\frac {(-1)^j}j Z(ijT-j\beta)=\sum_j{(-1)^j}j\int d\lambda \rho(\lambda) \exp(ijT\lambda-j\beta\lambda)
\end{equation}
In the case of the semicircle law, this becomes a sum of Bessel functions.

\begin{figure}
    \centering
    \includegraphics[scale=0.4]{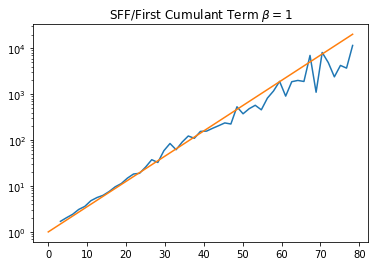}
    \includegraphics[scale=0.4]{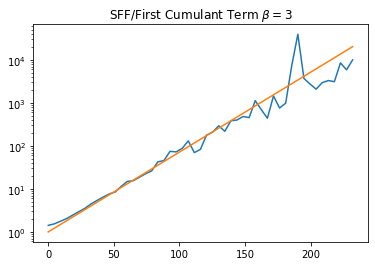}
    \includegraphics[scale=0.4]{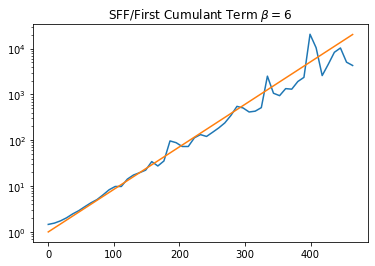}\\
    \includegraphics[scale=0.4]{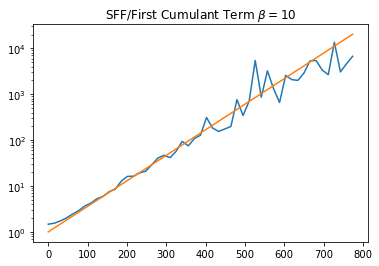}
    \includegraphics[scale=0.4]{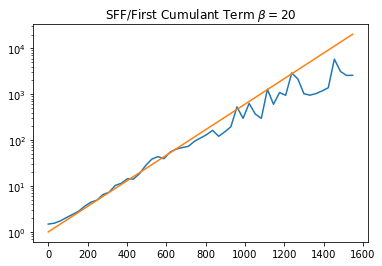}
    \includegraphics[scale=0.4]{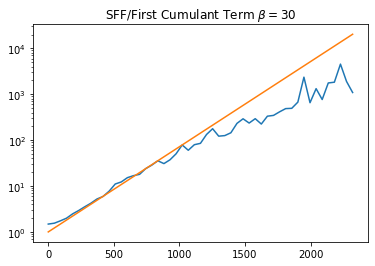}
    \caption{The quantity in blue is a numerical estimate for the SFF divided by the theoretical value from just the first cumulant. The value in orange is the second cumulant term, which seems to explain all the remaining discrepancy. At later times the two curves seem to diverge, suggesting that at sufficiently long times later terms in teh cumulant expansion might be necessary.}
    \label{fig:SYK2Graphs}
\end{figure}

The next term is 
\begin{equation}
    c_2=\textrm {var}\sum_{j\neq 0}{(-1)^j}jZ(ijT-j\beta)=\frac{1}{\bbeta\pi}\sum_{j\neq 0}\frac 1{j^2}\int d\lambda \exp(-2j\beta \lambda)jT
\end{equation}
The sum can be evaluated for any values of the parameters. In the case where $\beta E_{\text{max}}\gg1$, that sum becomes simple:
\begin{equation}
    \textrm {var}\sum_{j\neq 0}{(-1)^j}jZ(ijT-j\beta)=\frac{1}{\bbeta \pi}\sum_{j\neq 0} \frac{T}{2\beta j^2}=\frac{\pi T}{12\beta}
\end{equation}
We can graph $\log \left(\text{SFF}(T,\beta)\right)-c_1$ and $c_2$ as functions of time. We see good alignment in figure \ref{fig:SYK2Graphs}.

These findings can be interestingly compared with the so-called Sigma model, one of the state-of-the-art tools in spectral statistics. This technique consists of studying a many-body version of a one-body chaotic system \cite{Altland_2021,barney2023spectral,Altland_2021Operator}. By calculating the partition function of a Sigma Model, one can extract the spectral determinant, and thus the SFF, of the system of interest. The work in this paper as well as \cite{PhysRevLett.125.250601,Winer_2020} shows that not just the partition function but also the SFF of the many-body system contains signatures of the single-body SFF.

\section{Nonlinear Hydrodynamics and Interacting Sound Waves}
\label{sec:Interaction}

In this appendix, we sketch the inclusion of hydrodynamic interactions using a diagrammatic perturbation theory along the lines discussed in~\cite{WinerHydro}. We consider the basic diagramatic setup and, as an example, evaluate a single diagram. We do not give a comprehensive analysis of interaction effects and we discuss some key open issues at the end of the section.

Let's modify our action in equation \eqref{eq:lhydro1} to include interactions among the various modes. We now have a Lagrangian that includes a huge number of possible cubic and higher-order interactions. This leads to perturbative corrections to $\log \Zs$. For concreteness, we will choose the Lagrangian
\begin{equation}
        L_{\text{Kel}}=\frac 12 \phi_a\left(\partial_t^2+\frac{2\Gamma}{c^2} \partial_t^3-c^2\partial_\mu^2\right)\phi_r+\frac{2i\Gamma}{\beta c^2}\phi_a\partial_t^2\phi_a+\lambda(\partial_\mu \phi_r)^2 \partial_t \phi_r \partial_t \phi_a,
\end{equation}
though there are other quadratic terms with comparable effects, and the action we wrote isn't even KMS invariant.

We evaluate corrections to $\Zs$ diagrammatically, according to the rules in \cite{WinerHydro}. The propagators in the CTP formalism are
\begin{equation}
    \begin{split}
        G^{CTP}_{ra}(t,k)=\textrm{Re} \frac {e^{-\Gamma k^2t}}{ic|k|-\Gamma k^2} \exp\left(ic|k|t\right)\theta(t),\\
        G^{CTP}_{rr}(t,k)=\textrm{Re} \frac {e^{-\Gamma k^2|t|}}{ic|k|-\Gamma k^2} \exp\left(ic|k||t|\right).
    \end{split}
\end{equation}
In the DPT formalism, propagators are wrapped around the time circle according to
\begin{equation}
    G^{DPT}(t,k)=\sum_{n=-\infty}^\infty G^{CTP}(t+nT,k).
    \label{eq:GDPT}
\end{equation}
Note that the IR divergence for a sound-pole system is much less pronounced than in the case of diffusive hydrodynamics.

\begin{figure}
    \centering
    \includegraphics[scale=0.3]{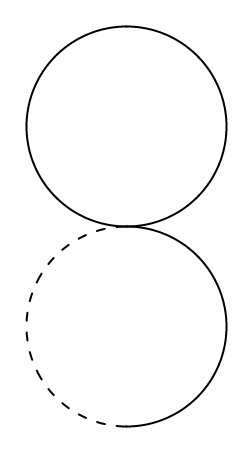}
    \caption{The dashed line represents $\phi_a$, the solid line represents $\phi_r$.}
    \label{fig:figureEight}
\end{figure}

With our propagators in hand, it is time to evaluate Feynman diagrams. Since the SFF is a partition function on an unorthodox manifold, the correction to the log of the SFF is given by a sum of bubble diagrams. The leading diagram is given in figure \ref{fig:figureEight}.
This can be factored to
\begin{equation}
    \Delta L=-\lambda\left(\int d^dk_1 k_1^2 G^{DPT}_{rr}(t=0,k_1)\right)\left(\int d^dk_2 \partial_t^2G^{DPT}_{ra}(t=0,k_2)\right) 
    \label{eq:2ints}
\end{equation}
The first of the integrals is UV divergent. We will impose a hard cutoff at $k=\Lambda$. In keeping with realistic hydrodynamics, we will chose $\Lambda\ll \frac c\Gamma$. With this cutoff, the first integral works out to $\frac{S_{d-1}}{d+2}\frac{\Gamma}{c^2}\Lambda^{d+2}$.
The second integral can be evaluated as
\begin{equation}
\begin{split}
\int d^dk_2 \partial_t^2G^{DPT}_{ra}(t=0,k)=\sum_{j=1}^\infty \int d^dk \partial_t^2G^{CTP}_{ra}(jT,k)=
\textrm{Re}\sum_{j=1}^\infty \int d^dk (ic|k|-\Gamma k^2){e^{-j\Gamma k^2T}}\exp\left(ijc|k|T\right)=\\
S_{d-1}\textrm{Re}\sum_{j=1}^\infty \int_0^\infty dk (ick-\Gamma k^2)|k|^{d-1}{e^{-j\Gamma k^2T}}\exp\left(ijckT\right)=\\
\frac{S_{d-1}}{2}\sum_{j=1}^\infty \int_{-\infty}^\infty dk (ick-\Gamma k^2)|k|^{d-1}{e^{-j\Gamma k^2T}}\exp\left(ijckT\right)
\end{split}
\label{eq:GraSum}
\end{equation}
Much like equation \eqref{eq:dip}, the integral in equation \eqref{eq:GraSum} has long-time behavior if $d$ is even, and a quick decay if $d$ is odd. The integral works out to
\begin{equation}
    \int_{-\infty}^\infty dk (ick-\Gamma k^2)|k|^{d-1}{e^{-j\Gamma k^2T}}\exp\left(ijckT\right)= 
\begin{cases}
  \sqrt{\frac{2\pi }{j\Gamma T}}\left(c\frac 1{(jT)^d}\partial_c^d-\Gamma\frac 1{(jT)^{d+1}}\partial_c^{d+1}\right)\exp(-j\frac{c^2}{4\Gamma}T)&\text{if }d \textrm{ odd}\\
  c\frac{(d)!}{(ijcT)^{d+1}}+O(T^{-d-2})          &\text{if }d \textrm{ even}
\end{cases}
\label{eq:cases}
\end{equation}
For odd $d$, the sum in equation \eqref{eq:GraSum} is dominated by the $j=1$ term, while for even $d$ we extract a value of $c\zeta(d+1)\frac{(d)!}{(icT)^{d+1}}+O(T^{-d-2})$.

Multiplying everything together, we get
\begin{equation}
    \Delta \text{(action)}/(VT)\approx\begin{cases}
        -\lambda \frac{S_{d-1}^2}{2(d+2)}\frac{\Gamma}{c^2}\Lambda^{d+2}\sqrt{\frac{2\pi }{\Gamma T}}\left(c\frac 1{(T)^d}\partial_c^d-\Gamma\frac 1{(T)^{d+1}}\partial_c^{d+1}\right)\exp(-\frac{c^2}{4\Gamma}T)&\text{if }d \textrm{ odd},\\
        -\lambda \frac{S_{d-1}^2}{2(d+2)}\frac{\Gamma}{c}\Lambda^{d+2}\zeta(d+1)\frac{(d)!}{(icT)^{d+1}}&\text{if }d \textrm{ even}.\\
    \end{cases}
\end{equation}
This can be interpreted as a renormalization of $c$ in equation \eqref{eq:dip}. More complicated diagrams have less straightforward interpretations, but any individual diagram can be evaluated using these Feynman rules. 

There are also more complicated effects with no obvious diagrammatic interpretation. All of the effects in this section are renormalizations to the first cumulant in the exponent. But presumably higher cumulants of the single-particle density of states can also be renormalized by interactions. We do not know how to systematically investigate these effects. For example, in writing equation \eqref{eq:2ints} we are assuming a density of states where the number of Laplacian eigenvalues between $k^2$ and $(k+\delta k)^2$ is $\frac{V}{(2\pi)^d}S_{d-1}k^{d-1}\delta k$. More realistically, this density would be a fluctuating variable. Would the overlaps between different eigenstates depend on the relative energies and on the density of states? We don't know. The higher-cumulant analogues of the calculation in this appendix are an open question.
\bibliographystyle{jhep}
\bibliography{bibFile.bib}

\providecommand{\href}[2]{#2}\begingroup\raggedright\begin{thebibliography}{10}

\bibitem{haake2010quantum}
F.~Haake, \emph{Quantum Signatures of Chaos}.
\newblock Springer Series in Synergetics. Springer Berlin Heidelberg, 2010.

\bibitem{PhysRevLett.52.1}
O.~Bohigas, M.~J. Giannoni and C.~Schmit, \emph{Characterization of chaotic
  quantum spectra and universality of level fluctuation laws},
  \href{http://dx.doi.org/10.1103/PhysRevLett.52.1}{\emph{Phys. Rev. Lett.}
  {\bf 52} (Jan, 1984) 1--4}.

\bibitem{mehta2004random}
M.~Mehta, \emph{Random Matrices}.
\newblock ISSN. Elsevier Science, 2004.

\bibitem{doi:10.1063/1.1703775}
F.~J. Dyson, \emph{Statistical theory of the energy levels of complex systems.
  iii}, \href{http://dx.doi.org/10.1063/1.1703775}{\emph{Journal of
  Mathematical Physics} {\bf 3} (1962) 166--175},
  [\href{https://arxiv.org/abs/https://doi.org/10.1063/1.1703775}{{\tt
  https://doi.org/10.1063/1.1703775}}].

\bibitem{wigner1959group}
E.~Wigner and J.~Griffin, \emph{Group Theory and Its Application to the Quantum
  Mechanics of Atomic Spectra}.
\newblock Pure and applied Physics. Academic Press, 1959.

\bibitem{berry1977level}
M.~V. Berry and M.~Tabor, \emph{Level clustering in the regular spectrum},
  {\emph{Proceedings of the Royal Society of London. A. Mathematical and
  Physical Sciences} {\bf 356} (1977) 375--394}.

\bibitem{McDonald}
S.~W. McDonald and A.~N. Kaufman, \emph{Spectrum and eigenfunctions for a
  hamiltonian with stochastic trajectories},
  \href{http://dx.doi.org/10.1103/PhysRevLett.42.1189}{\emph{Phys. Rev. Lett.}
  {\bf 42} (Apr, 1979) 1189--1191}.

\bibitem{BERRY1981163}
M.~Berry, \emph{Quantizing a classically ergodic system: Sinai's billiard and
  the kkr method},
  \href{http://dx.doi.org/https://doi.org/10.1016/0003-4916(81)90189-5}{\emph{Annals
  of Physics} {\bf 131} (1981) 163--216}.

\bibitem{bohigas1984chaotic}
O.~Bohigas and M.-J. Giannoni, \emph{Chaotic motion and random matrix
  theories},  in \emph{Mathematical and computational methods in nuclear
  physics}, pp.~1--99.
\newblock Springer, 1984.

\bibitem{altshuler1986repulsion}
B.~L. Altshuler and B.~I. Shklovskii, \emph{Repulsion of energy levels and
  conductivity of small metal samples}, {\emph{Sov. Phys. JETP} {\bf 64} (1986)
  127}.

\bibitem{andreev1995spectral}
A.~V. Andreev and B.~L. Altshuler, \emph{Spectral statistics beyond random
  matrix theory}, {\emph{Phys. Rev. Lett.} {\bf 75} (Jul, 1995) 902--905}.

\bibitem{dubertrand2016spectral}
R.~Dubertrand and S.~M{\"u}ller, \emph{Spectral statistics of chaotic many-body
  systems}, {\emph{New Journal of Physics} {\bf 18} (2016) 033009}.

\bibitem{PhysRevLett.121.264101}
B.~Bertini, P.~Kos and T.~c.~v. Prosen, \emph{Exact spectral form factor in a
  minimal model of many-body quantum chaos},
  \href{http://dx.doi.org/10.1103/PhysRevLett.121.264101}{\emph{Phys. Rev.
  Lett.} {\bf 121} (Dec, 2018) 264101}.

\bibitem{Wittmann_W__2022}
K.~W. W., E.~R. Castro, A.~Foerster and L.~F. Santos, \emph{Interacting bosons
  in a triple well: Preface of many-body quantum chaos},
  \href{http://dx.doi.org/10.1103/physreve.105.034204}{\emph{Physical Review E}
  {\bf 105} (mar, 2022) }.

\bibitem{speck}
L.~Santos, F.~Perez-Bernal and E.~Torres-Herrera, \emph{Speck of chaos},
  \href{http://dx.doi.org/10.1103/PhysRevResearch.2.043034}{\emph{Physical
  Review Research} {\bf 2} (09, 2020) 43034}.

\bibitem{bunin2023fisher}
G.~Bunin, L.~Foini and J.~Kurchan, \emph{Fisher zeroes and the fluctuations of
  the spectral form factor of chaotic systems},  2023.

\bibitem{kamenev_1999}
A.~Kamenev and M.~M{\'{e} }zard, \emph{Wigner-dyson statistics from the replica
  method}, \href{http://dx.doi.org/10.1088/0305-4470/32/24/304}{\emph{Journal
  of Physics A: Mathematical and General} {\bf 32} (jan, 1999) 4373--4388}.

\bibitem{PhysRevResearch.3.L012019}
A.~Prakash, J.~H. Pixley and M.~Kulkarni, \emph{Universal spectral form factor
  for many-body localization},
  \href{http://dx.doi.org/10.1103/PhysRevResearch.3.L012019}{\emph{Phys. Rev.
  Res.} {\bf 3} (Feb, 2021) L012019}.

\bibitem{Cotler2017}
J.~S. Cotler, G.~Gur-Ari, M.~Hanada, J.~Polchinski, P.~Saad, S.~H. Shenker
  et~al., \emph{Black holes and random matrices},
  \href{http://dx.doi.org/10.1007/jhep05(2017)118}{\emph{Journal of High Energy
  Physics} {\bf 2017} (May, 2017) }.

\bibitem{Saad:2018bqo}
P.~Saad, S.~H. Shenker and D.~Stanford, \emph{{A semiclassical ramp in SYK and
  in gravity}},  \href{https://arxiv.org/abs/1806.06840}{{\tt 1806.06840}}.

\bibitem{PhysRevD.98.086026}
J.~Liu, \emph{Spectral form factors and late time quantum chaos},
  \href{http://dx.doi.org/10.1103/PhysRevD.98.086026}{\emph{Phys. Rev. D} {\bf
  98} (Oct, 2018) 086026}.

\bibitem{saad2019late}
P.~Saad, \emph{Late time correlation functions, baby universes, and eth in jt
  gravity},  2019.

\bibitem{saad2021wormholes}
P.~Saad, S.~H. Shenker, D.~Stanford and S.~Yao, \emph{Wormholes without
  averaging},  2021.

\bibitem{saad2022convergent}
P.~Saad, D.~Stanford, Z.~Yang and S.~Yao, \emph{A convergent genus expansion
  for the plateau},  2022.

\bibitem{shivam_2023}
S.~Shivam, A.~D. Luca, D.~A. Huse and A.~Chan, \emph{Many-body quantum chaos
  and emergence of ginibre ensemble},
  \href{http://dx.doi.org/10.1103/physrevlett.130.140403}{\emph{Physical Review
  Letters} {\bf 130} (apr, 2023) }.

\bibitem{Chan_2018}
A.~Chan, A.~D. Luca and J.~Chalker, \emph{Spectral statistics in spatially
  extended chaotic quantum many-body systems},
  \href{http://dx.doi.org/10.1103/physrevlett.121.060601}{\emph{Physical Review
  Letters} {\bf 121} (aug, 2018) }.

\bibitem{Chan_2018Min}
A.~Chan, A.~D. Luca and J.~Chalker, \emph{Solution of a minimal model for
  many-body quantum chaos},
  \href{http://dx.doi.org/10.1103/physrevx.8.041019}{\emph{Physical Review X}
  {\bf 8} (nov, 2018) }.

\bibitem{Chan_2019}
A.~Chan, A.~D. Luca and J.~T. Chalker, \emph{Eigenstate correlations,
  thermalization, and the butterfly effect},
  \href{http://dx.doi.org/10.1103/physrevlett.122.220601}{\emph{Physical Review
  Letters} {\bf 122} (jun, 2019) }.

\bibitem{Liao_2022}
Y.~Liao and V.~Galitski, \emph{Effective field theory of random quantum
  circuits}, \href{http://dx.doi.org/10.3390/e24060823}{\emph{Entropy} {\bf 24}
  (jun, 2022) 823}.

\bibitem{PhysRevLett.127.230602}
H.~Singh, B.~A. Ware, R.~Vasseur and A.~J. Friedman, \emph{Subdiffusion and
  many-body quantum chaos with kinetic constraints},
  \href{http://dx.doi.org/10.1103/PhysRevLett.127.230602}{\emph{Phys. Rev.
  Lett.} {\bf 127} (Dec, 2021) 230602}.

\bibitem{CCRM}
J.~Cotler, N.~Hunter-Jones, J.~Liu and B.~Yoshida, \emph{Chaos, complexity, and
  random matrices},
  \href{http://dx.doi.org/10.1007/jhep11(2017)048}{\emph{Journal of High Energy
  Physics} {\bf 2017} (nov, 2017) }.

\bibitem{Kos_2018}
P.~Kos, M.~Ljubotina and T.~Prosen, \emph{Many-body quantum chaos: Analytic
  connection to random matrix theory},
  \href{http://dx.doi.org/10.1103/physrevx.8.021062}{\emph{Physical Review X}
  {\bf 8} (jun, 2018) }.

\bibitem{Flack_2020}
A.~Flack, B.~Bertini and T.~Prosen, \emph{Statistics of the spectral form
  factor in the self-dual kicked ising model},
  \href{http://dx.doi.org/10.1103/physrevresearch.2.043403}{\emph{Physical
  Review Research} {\bf 2} (dec, 2020) }.

\bibitem{2020Prosen}
D.~Roy and T.~Prosen, \emph{Random matrix spectral form factor in kicked
  interacting fermionic chains},
  \href{http://dx.doi.org/10.1103/physreve.102.060202}{\emph{Physical Review E}
  {\bf 102} (Dec, 2020) }.

\bibitem{PhysRevLett.125.250601}
Y.~Liao, A.~Vikram and V.~Galitski, \emph{Many-body level statistics of
  single-particle quantum chaos},
  \href{http://dx.doi.org/10.1103/PhysRevLett.125.250601}{\emph{Phys. Rev.
  Lett.} {\bf 125} (Dec, 2020) 250601}.

\bibitem{Winer_2020}
M.~Winer, S.-K. Jian and B.~Swingle, \emph{Exponential ramp in the quadratic
  sachdev-ye-kitaev model},
  \href{http://dx.doi.org/10.1103/physrevlett.125.250602}{\emph{Physical Review
  Letters} {\bf 125} (dec, 2020) }.

\bibitem{Li_2021}
J.~Li, T.~Prosen and A.~Chan, \emph{Spectral statistics of non-hermitian
  matrices and dissipative quantum chaos},
  \href{http://dx.doi.org/10.1103/physrevlett.127.170602}{\emph{Physical Review
  Letters} {\bf 127} (oct, 2021) }.

\bibitem{Joshi_2022}
L.~K. Joshi, A.~Elben, A.~Vikram, B.~Vermersch, V.~Galitski and P.~Zoller,
  \emph{Probing many-body quantum chaos with quantum simulators},
  \href{http://dx.doi.org/10.1103/physrevx.12.011018}{\emph{Physical Review X}
  {\bf 12} (jan, 2022) }.

\bibitem{ma2022quantum}
C.-T. Ma and C.-H. Wu, \emph{Quantum entanglement and spectral form factor},
  2022.

\bibitem{2021}
A.~Ahmed, N.~Roy and A.~Sharma, \emph{Dynamics of spectral correlations in the
  entanglement hamiltonian of the aubry-andré-harper model},
  \href{http://dx.doi.org/10.1103/physrevb.104.155137}{\emph{Physical Review B}
  {\bf 104} (Oct, 2021) }.

\bibitem{shaffer_2014}
D.~Shaffer, C.~Chamon, A.~Hamma and E.~R. Mucciolo, \emph{Irreversibility and
  entanglement spectrum statistics in quantum circuits},
  \href{http://dx.doi.org/10.1088/1742-5468/2014/12/p12007}{\emph{Journal of
  Statistical Mechanics: Theory and Experiment} {\bf 2014} (dec, 2014) P12007}.

\bibitem{WinerLoschmidt}
M.~Winer and B.~Swingle, \emph{The loschmidt spectral form factor},
  \href{http://dx.doi.org/10.1007/jhep10(2022)137}{\emph{Journal of High Energy
  Physics} {\bf 2022} (oct, 2022) }.

\bibitem{Weidenmuller_2005}
H.~A. Weidenmüller, \emph{Parametric level correlations in random-matrix
  models}, \href{http://dx.doi.org/10.1088/0953-8984/17/20/015}{\emph{Journal
  of Physics: Condensed Matter} {\bf 17} (may, 2005) S1881--S1887}.

\bibitem{guhr1998random}
T.~Guhr, A.~M{\"u}ller-Groeling and H.~A. Weidenm{\"u}ller, \emph{Random-matrix
  theories in quantum physics: common concepts}, {\emph{Physics Reports} {\bf
  299} (1998) 189--425}.

\bibitem{https://doi.org/10.48550/arxiv.2205.12968}
J.~Cotler and K.~Jensen, \emph{A precision test of averaging in ads/cft},
  2022.
\newblock 10.48550/ARXIV.2205.12968.

\bibitem{PhysRevLett.70.4063}
B.~D. Simons and B.~L. Altshuler, \emph{Universal velocity correlations in
  disordered and chaotic systems},
  \href{http://dx.doi.org/10.1103/PhysRevLett.70.4063}{\emph{Phys. Rev. Lett.}
  {\bf 70} (Jun, 1993) 4063--4066}.

\bibitem{WinerHydro}
M.~Winer and B.~Swingle, \emph{Hydrodynamic theory of the connected spectral
  form factor},  2020.
\newblock 10.48550/ARXIV.2012.01436.

\bibitem{Winer_2022}
M.~Winer and B.~Swingle, \emph{Spontaneous symmetry breaking, spectral
  statistics, and the ramp},
  \href{http://dx.doi.org/10.1103/physrevb.105.104509}{\emph{Physical Review B}
  {\bf 105} (mar, 2022) }.

\bibitem{Friedman_2019}
A.~J. Friedman, A.~Chan, A.~De~Luca and J.~Chalker, \emph{Spectral statistics
  and many-body quantum chaos with conserved charge},
  \href{http://dx.doi.org/10.1103/physrevlett.123.210603}{\emph{Physical Review
  Letters} {\bf 123} (Nov, 2019) }.

\bibitem{moudgalya2020spectral}
S.~Moudgalya, A.~Prem, D.~A. Huse and A.~Chan, \emph{Spectral statistics in
  constrained many-body quantum chaotic systems},  2020.

\bibitem{kos_2021}
P.~Kos, T.~Prosen and B.~Bertini, \emph{Thermalization dynamics and spectral
  statistics of extended systems with thermalizing boundaries},
  \href{http://dx.doi.org/10.1103/physrevb.104.214303}{\emph{Physical Review B}
  {\bf 104} (dec, 2021) }.

\bibitem{spohn_2020}
H.~Spohn, \emph{The 1+1 dimensional kardar–parisi–zhang equation: more
  surprises}, .

\bibitem{KPZ}
M.~Kardar, G.~Parisi and Y.-C. Zhang, \emph{Dynamic scaling of growing
  interfaces}, \href{http://dx.doi.org/10.1103/PhysRevLett.56.889}{\emph{Phys.
  Rev. Lett.} {\bf 56} (Mar, 1986) 889--892}.

\bibitem{Krug1997OriginsOS}
J.~H.~A. Krug, \emph{Origins of scale invariance in growth processes},
  {\emph{Advances in Physics} {\bf 46} (1997) 139--282}.

\bibitem{krajnik_2020}
{\v{Z}}.~Krajnik and T.~Prosen,
  \emph{Kardar{\textendash}parisi{\textendash}zhang physics in integrable
  rotationally symmetric dynamics on discrete space{\textendash}time lattice},
  \href{http://dx.doi.org/10.1007/s10955-020-02523-1}{\emph{Journal of
  Statistical Physics} {\bf 179} (mar, 2020) 110--130}.

\bibitem{Altland1997}
A.~Altland and M.~R. Zirnbauer, \emph{Nonstandard symmetry classes in
  mesoscopic normal-superconducting hybrid structures},
  \href{http://dx.doi.org/10.1103/PhysRevB.55.1142}{\emph{Phys. Rev. B} {\bf
  55} (Jan, 1997) 1142--1161}.

\bibitem{tao2012topics}
T.~Tao, \emph{{Topics in Random Matrix Theory}}.
\newblock Graduate studies in mathematics. American Mathematical Society, 2012.

\bibitem{Gharibyan_2018}
H.~Gharibyan, M.~Hanada, S.~H. Shenker and M.~Tezuka, \emph{Onset of random
  matrix behavior in scrambling systems},
  \href{http://dx.doi.org/10.1007/jhep07(2018)124}{\emph{Journal of High Energy
  Physics} {\bf 2018} (jul, 2018) }.

\bibitem{glorioso2018lectures}
H.~Liu and P.~Glorioso, \emph{{Lectures on non-equilibrium effective field
  theories and fluctuating hydrodynamics}},
  \href{http://dx.doi.org/10.22323/1.305.0008}{\emph{PoS} {\bf TASI2017} (2018)
  008}, [\href{https://arxiv.org/abs/1805.09331}{{\tt 1805.09331}}].

\bibitem{crossley2017effective}
M.~Crossley, P.~Glorioso and H.~Liu, \emph{{Effective field theory of
  dissipative fluids}},
  \href{http://dx.doi.org/10.1007/JHEP09(2017)095}{\emph{JHEP} {\bf 09} (2017)
  095}, [\href{https://arxiv.org/abs/1511.03646}{{\tt 1511.03646}}].

\bibitem{Glorioso_2017}
P.~Glorioso, M.~Crossley and H.~Liu, \emph{Effective field theory of
  dissipative fluids (ii): classical limit, dynamical kms symmetry and entropy
  current}, \href{http://dx.doi.org/10.1007/jhep09(2017)096}{\emph{Journal of
  High Energy Physics} {\bf 2017} (Sep, 2017) 1--44}.

\bibitem{gao2018ghostbusters}
P.~Gao, P.~Glorioso and H.~Liu, \emph{{Ghostbusters: Unitarity and Causality of
  Non-equilibrium Effective Field Theories}},
  \href{http://dx.doi.org/10.1007/JHEP03(2020)040}{\emph{JHEP} {\bf 03} (2020)
  040}, [\href{https://arxiv.org/abs/1803.10778}{{\tt 1803.10778}}].

\bibitem{Grozdanov_2015}
S.~Grozdanov and J.~Polonyi, \emph{Viscosity and dissipative hydrodynamics from
  effective field theory},
  \href{http://dx.doi.org/10.1103/physrevd.91.105031}{\emph{Physical Review D}
  {\bf 91} (May, 2015) }.

\bibitem{Kovtun_2012}
P.~Kovtun, \emph{Lectures on hydrodynamic fluctuations in relativistic
  theories},
  \href{http://dx.doi.org/10.1088/1751-8113/45/47/473001}{\emph{Journal of
  Physics A: Mathematical and Theoretical} {\bf 45} (Nov, 2012) 473001}.

\bibitem{Dubovsky_2012}
S.~Dubovsky, L.~Hui, A.~Nicolis and D.~T. Son, \emph{Effective field theory for
  hydrodynamics: Thermodynamics, and the derivative expansion},
  \href{http://dx.doi.org/10.1103/physrevd.85.085029}{\emph{Physical Review D}
  {\bf 85} (Apr, 2012) }.

\bibitem{Endlich_2013}
S.~Endlich, A.~Nicolis, R.~A. Porto and J.~Wang, \emph{Dissipation in the
  effective field theory for hydrodynamics: First-order effects},
  \href{http://dx.doi.org/10.1103/physrevd.88.105001}{\emph{Physical Review D}
  {\bf 88} (Nov, 2013) }.

\bibitem{Chan_2022}
A.~Chan, S.~Shivam, D.~A. Huse and A.~D. Luca, \emph{Many-body quantum chaos
  and space-time translational invariance},
  \href{http://dx.doi.org/10.1038/s41467-022-34318-1}{\emph{Nature
  Communications} {\bf 13} (dec, 2022) }.

\bibitem{Freivogel_2012}
B.~Freivogel, J.~McGreevy and S.~J. Suh, \emph{Exactly stable collective
  oscillations in conformal field theory},
  \href{http://dx.doi.org/10.1103/physrevd.85.105002}{\emph{Physical Review D}
  {\bf 85} (may, 2012) }.

\bibitem{luca}
L.~V. Delacretaz, A.~L. Fitzpatrick, E.~Katz and M.~T. Walters,
  \emph{Thermalization and chaos in a 1+1d qft},  2022.
\newblock 10.48550/ARXIV.2207.11261.

\bibitem{Kravtsov1994LevelSI}
V.~E. Kravtsov and A.~D. Mirlin, \emph{Level statistics in a metallic sample:
  Corrections to the wigner-dyson distribution}, {\emph{arXiv: Condensed
  Matter} (1994) }.

\bibitem{Altland_2021}
A.~Altland and J.~Sonner, \emph{Late time physics of holographic quantum
  chaos},
  \href{http://dx.doi.org/10.21468/scipostphys.11.2.034}{\emph{{SciPost}
  Physics} {\bf 11} (aug, 2021) }.

\bibitem{barney2023spectral}
R.~Barney, M.~Winer, C.~L. Baldwin, B.~Swingle and V.~Galitski, \emph{Spectral
  statistics of a minimal quantum glass model},  2023.

\bibitem{Altland_2021Operator}
A.~Altland, D.~Bagrets, P.~Nayak, J.~Sonner and M.~Vielma, \emph{From operator
  statistics to wormholes},
  \href{http://dx.doi.org/10.1103/physrevresearch.3.033259}{\emph{Physical
  Review Research} {\bf 3} (sep, 2021) }.

\end{thebibliography}\endgroup
\end{document}